\documentclass[twocolumn]{svjour3}          

\smartqed  
\usepackage{graphicx}
\usepackage{caption}
\usepackage{multirow}
\usepackage{array}
\usepackage{amsmath}
\usepackage{subfigure}
\usepackage{float}
\usepackage{lipsum}
\usepackage[numbers,sort&compress]{natbib}
\setcitestyle{square, comma, numbers, sort&compress}
\usepackage{fix-cm}
\usepackage{enumitem}
\usepackage{placeins} 
\graphicspath{{./Figures/}}
\begin{document}\sloppy

\title{Explainable COVID-19 Detection Using Chest CT Scans\\ and Deep Learning}

\author{Hammam Alshazly$^{1,2}$ \and Christoph Linse$^{1}$ \and Erhardt Barth$^{1}$ \and Thomas Martinetz$^{1}$}

\authorrunning{Alshazly et al.} 

\institute{ 
	$^1$Institute for Neuro- and Bioinformatics, University of L{\"u}beck, 23562 L{\"u}beck, Germany\\		
			$^2$Department of Mathematics, Faculty of Science, South Valley University,  Qena 83523, Egypt\\
			}
\date{Received: date / Accepted: date}

\maketitle

\begin{abstract}
	
This paper explores how well deep learning models trained on chest CT images can diagnose COVID-19 infected people in a fast and automated process. To this end, we adopt advanced deep network architectures and propose a transfer learning strategy using custom-sized input tailored for each deep architecture to achieve the best  performance. We conduct extensive sets of experiments on two CT image datasets, namely the SARS-CoV-2 CT-scan and the COVID19-CT. The obtained results show superior performances for our models compared with previous studies, where our best models achieve average accuracy, precision, sensitivity, specificity and F1 score of $99.4\%$, $99.6\%$, $99.8\%$, $99.6\%$ and $99.4\%$ on the SARS-CoV-2 dataset; and $92.9\%$, $91.3\%$, $93.7\%$, $92.2\%$ and $92.5\%$ on the COVID19-CT dataset, respectively. Furthermore, we apply two visualization techniques to provide visual explanations for the models' predictions. The visualizations show well-separated clusters for CT images of COVID-19 from other lung diseases, and accurate localizations of the COVID-19 associated regions.


\end{abstract}
\vspace{-0.5 cm}
\section{Introduction}
\label{introduction}
Coronavirus disease 2019 (COVID-19) is an infectious disease caused by the new coronavirus named severe acute respiratory syndrome coronavirus-2 (SARS-CoV-2). The virus is highly contagious and can be transmitted by direct and/or indirect contact with infected people through respiratory droplets when they sneeze, cough or even talk~\cite{liu2020community, ghinai2020first, chen2020epidemiological}. The real-time polymerase chain reaction (RT-PCR) test is the standard reference for confirming COVID-19, and with the rapid increment in the number of infected people, most of the countries are encountering shortage in testing kits. Moreover, RT-PCR testing has high turnaround times and a high false negative rate \cite{long2020diagnosis}. Thus, it is highly desirable to consider other testing tools for identifying COVID-19 contaminated patients to isolate them and mitigate the pandemic impact on the life of many people. 

Chest computed tomography (CT) is an applicable supplement to RT-PCR testing and has been playing a role in screening and diagnosing COVID-19 infections. In recent studies~\cite{fang2020sensitivity, ai2020correlation}, the authors manually examined chest CT scans for more than a thousand patients and confirmed the usefulness of chest CT scans for diagnosing COVID-19 with high sensitivity rates. In some cases, the patients initially had a negative PCR test, however, confirmation was based on their positive CT findings. Moreover, chest CT screening has been recommended when patients show symptoms compatible with viral infections, but the result of their PCR test is negative \cite{fang2020sensitivity, kanne2020chest}. Nevertheless, diagnosing COVID-19 from chest CT images by radiologists takes time, and manually checking every CT image might not be feasible in emergency cases. Therefore, there is a need for automated detection tools that exploit the recent deep learning techniques and CT images to expedite the process and provide consistent performance. 

This paper adopts the most advanced deep Convolutional Neural Network (CNN) architectures, which are top performers in the ImageNet recognition challenge~\cite{russakovsky2015imagenet}, and presents a comprehensive study for detecting COVID-19 based on CT images. We explore CNN models that have different architectural designs and varying depths to obtain the best detection performance. Even though we conduct our experiments on two of the largest CT scan datasets available for research, their size is still insufficient to train deep networks from scratch. An effective strategy to overcome this limitation is to use transfer learning \cite{weiss2016survey}, where deep networks trained on visual tasks are utilized to initialize networks for different but related target tasks. Most of the published works that applied transfer learning strategies using the ImageNet \cite{deng2009imagenet} pretrained networks followed the strict fixed-sized input for each deep network and resized their target images accordingly. We argue that resizing images with different aspect ratios to match a specific resolution can distort the image severely. We address the problem by placing the images into a fixed-sized canvas determined specifically for each CNN architecture, where the aspect ration of the original image is preserved. This has proven to be a less violating procedure and more effective to achieve better results as reported in \cite{alshazly2020deep}. Moreover, we utilize the layer-wise adaptive large batch optimization technique called LAMB \cite{you2020large}, which has demonstrated better performance and convergence speed for training deep networks. The performance of the models is measured quantitatively using accuracy, precision, sensitivity, specificity, F1-score and the confusion matrix for each model. Our obtained results indicate the effectiveness of our strategy to achieve state-of-the-art results on the considered datasets. 

In order to provide better explainability of the deep models and making them more transparent we apply two visualization techniques. The first approach is the t-distributed Stochastic Neighboring Embedding (t-SNE)~\cite{maaten2008visualizing}, which is a dimensionality reduction and visualization technique for visualizing clusters of instances in a high-dimensional space. The obtained visualizations of the t-SNE embeddings show well-separated clusters representing CT images for COVID-19 and Non-COVID-19 cases. The second approach is the Gradient-weighted Class Activation Mapping (Grad-CAM)~\cite{selvaraju2017grad}, which is a visualization technique for CNN-based models. It provides high-resolution and class-discriminative visualizations that localize the important image regions considered for the model prediction. The Grad-CAM visualizations show how accurately our models localize the COVID-19 associated regions. Overall, this paper exhibits the following contributions:

\begin{itemize}
	
	\item A comparative experimental study is conducted on how well advanced deep CNNs trained on chest CT images can identify COVID-19 cases. To this end, we experiment with 12 deep networks that have different architectural designs and varying depths, and provide quantitative and qualitative analyses.\\ 
	
	\item We propose a domain adaptation strategy to fine-tune deep networks using custom-sized inputs determined specifically for each architecture, and utilize the LAMB optimizer for training the networks. Our experimental results prove the effectiveness of our optimization configurations to obtain state-of-the-art performance on the considered CT image datasets. Our best models achieve an average accuracy of $99.4\%$ and $92.9\%$, and average sensitivity rates of $99.8\%$ and $93.7\%$ on the largest datasets of CT images available for research.\\ 
	
	\item We provide visualizations of the extracted features from different models to understand how deep networks represent CT images in the feature space. The visualizations show well-separated clusters representing the CT images of the different classes, which indicates that our models have learned discriminative features to distinguish CT images of different cases.\\
	
	\item We show discriminative localizations and visual explanations obtained by our models for detecting COVID-19 associated regions in CT images as annotated by expert radiologists. 
	
\end{itemize}
The rest of the paper is structured as follows. We review the related work in the next section. The deep CNN architectures are described in Section~\ref{deep_networks} and the methodology to learn discriminative features in Section~\ref{methodology}. The experimental settings and the obtained results are reported in Section~\ref{experiments}. Finally, we draw the main conclusion in Section~\ref{conclusion}.
\vspace{-0.5 cm}
\section{Related Work}
\label{related_work}
This section highlights some relevant work that adopted deep CNNs for building computer-aided diagnosis (CADs) systems based on medical images. The authors in \cite{shin2016deep} employed different deep CNN architectures, which were pretrained on the ImageNet dataset \cite{deng2009imagenet}, and fine-tuned them on specific CT scans for thoraco-abdominal lymph node detection and interstitial lung disease classification. Their study indicated the effectiveness of deep CNNs for CADs problems even when training data is limited. In \cite{rajpurkar2017chexnet}, the authors proposed the CheXNet model to detect different types of pneumonia from chest X-ray images. The model consisted of 121-layers and was trained on a large dataset that contained over 100,000 X-ray images for 14 different thoracic diseases. The model showed outstanding detection performance at the level of practicing radiologists.  

In the context of the COVID-19 pandemic, extensive research has been conducted to develop automated image-based COVID-19 detection and diagnosis systems \cite{li2020artificial, xu2020deep, zheng2020deep,wang2020deep, shan2020lung}. We hereafter review the proposed approaches for reliable detection systems based on chest X-ray and CT-scan imaging modalities. These techniques follow either one of two main paradigms. 

On one hand, new deep network architectures have been developed and tailored specifically for detecting and recognizing COVID-19. COVID-Net \cite{wang2020covid} represents one of the earliest convolutional networks designed for detecting COVID-19 cases automatically from X-ray images. The performance of the network showed an acceptable accuracy of $83.5\%$ and a high sensitivity of $100\%$ for COVID-19 cases. Hasan et al. \cite{hasan2020cvr} proposed a CNN-based network named Coronavirus Recognition Network (CVR-Net) to automatically detect COVID-19 cases from radiography images. The network was trained and evaluated on datasets with X-ray and CT images. The obtained results showed varying accuracy scores based on the number of classes in the underlying X-ray image dataset and an average accuracy of $78\%$ for the CT image dataset. Further modifications were applied to COVID-Net to improve its representational ability for one specific image modality and to make the network computationally more efficient as in \cite{wang2020contrastive}. 

On the other hand, some deep networks have been proposed for similar tasks of automated detection and recognition of COVID-19 cases, however, these networks are based on well-designed and existing CNN architectures, such as ReseNet \cite{farooq2020covid}, Xception \cite{khan2020coronet} and Capsule Networks \cite{afshar2020covid, toraman2020convolutional}. The authors in \cite{apostolopoulos2020covid} adopted transfer learning from deep networks for automatic COVID-19 detection based on X-ray images from patients with bacterial and COVID-19 pneumonia and normal cases. They reported the best results for the two- and three-class classification tasks with an accuracy of $98.75\%$ and $93.48$, respectively. Minaee et al. \cite{minaee2020deep} applied transfer learning by fine-tuning four popular pretrained CNNs to identify COVID-19 infection. They experimented on a prepared X-ray image dataset with 5,000 chest X-rays. Their best approaches obtained an average sensitivity and specificity of $98\%$ and $90\%$, respectively. Brunese et al. \cite{brunese2020explainable} utilized transfer learning with a pretrained VGG-16 network \cite{Simonyan2015very} to automatically detect COVID-19 from chest X-rays. On a combined dataset from different sources with X-rays for healthy and pulmonary disease they reported an average accuracy of $97\%$. 

Zhou et al. \cite{zhou2020improved} highlighted the importance of deep learning techniques and chest CT images for differentiating COVID-19 pneumonia and influenza pneumonia. The study was conducted on CT images for confirmed COVID-19 patients from different hospitals in china. Their study proved the potential of accurate COVID-19 diagnosis from CT images and the effectiveness of their proposed classification scheme to differentiate between the two types of pneumonia. DeepPneumonia \cite{song2020deep} was developed to identify COVID-19 cases (88 patients), bacterial pneumonia (100 patients) and healthy cases (86 subjects) based on CT images. The model achieved an accuracy of $86.5\%$ for differentiating bacterial and viral (COVID-19) pneumonia and an accuracy of $94\%$ for distinguishing COVID-19 and healthy cases. The authors in \cite{jaiswal2020classification} used CT images to classify COVID-19 infected patients from Non-COVID-19 people utilizing a pretrained DenseNet201 network. The model achieved an accuracy of $96.25\%$.

Very few studies employed handcrafted feature extraction methods and conventional classifiers. In \cite{pereira2020covid}, texture features were extracted from X-ray images using popular texture descriptors. The features were combined with those extracted from a pretrained Inception V3 \cite{inceptionnet2016} using different fusion strategies. Then, various classifiers were used to differentiate between normal X-rays and different types of pneumonia. The best classification scheme achieved an F1-score of $83\%$. In~\cite{al2020machine}, the authors proposed an approach to differentiate between positive and negative COVID-19 cases based on CT scans. Different texture features were extracted from CT images with Gabor filters, and then support vector machines were trained for classification. Their proposed scheme achieved an average accuracy of $95.37\%$ and a sensitivity of $95.99\%$.  

The discussion about related works indicates the prominence of deep learning methods to address the task of automated detection of COVID-19. We build on the existing body of published work and adopt advanced deep networks for detecting COVID-19 using CT images. We conduct experiments on two of the largest CT image datasets and compare the performance of 12 deep networks using standard evaluation metrics. We also provide visualizations for better explainability of the resulting models. 
\vspace{-0.5 cm}
\section{Deep Network Architectures} 
\label{deep_networks} 
This section describes the deep CNN architectures employed to identify COVID-19 using chest CT scans. These networks are state-of-the-art deep models for image recognition. They differ in their architectural design and were proposed in order to achieve better representational power or to reduce their computational complexity. In this work we consider the most advanced networks such as SqueezeNet~\cite{iandola2016squeezenet}, Inception~\cite{inceptionnet2016}, ResNet~\cite{he2016deep}, ResNeXt~\cite{xie2017aggregated}, Xception~\cite{chollet2017xception}, ShuffleNet~\cite{ma2018shufflenet} and DenseNet~\cite{densenet2017}. 

\subsection{SqueezeNet} 
\label{squeezenet} 
\begin{figure}[tp] 
	\centering 
	{\includegraphics[width=5 cm, height=6 cm]{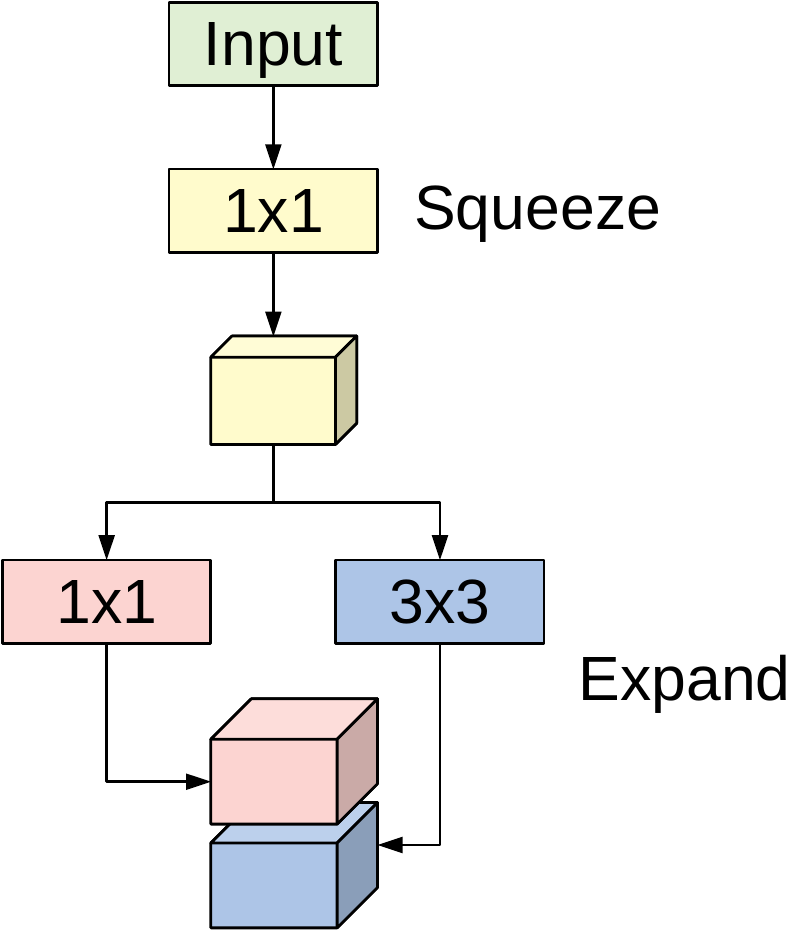}} 
	\caption{The fire module used in SqueezeNet.} 
	\label{fire_module} 
\end{figure}

The SqueezeNet architecture is a deep CNN proposed for computer vision tasks with the main concerns on efficiency (having fewer parameters and smaller model size)~\cite{iandola2016squeezenet}. The basic building block for the SqueezeNet architecture is the fire module depicted in Figure~\ref{fire_module}. The module incorporates the squeeze phase and the expand phase. The squeeze phase applies a set of $1 \times 1$ filters followed by a ReLU activation. The number of learned squeeze filters is always smaller than the size of the input volume. Consequently, the squeeze phase can be considered as a dimensionality reduction process, and at the same time it captures the pixel correlations across the input channels. The output of the squeeze phase is fed into the expand phase, in which a combination of $1 \times 1$ and $3 \times 3$ convolutions are learned. The larger $3 \times 3$ filters are used to capture the spatial correlation amongst pixels. The outputs of the expand phase are concatenated across the channel dimension and then evaluated by a ReLU activation. 

The original paper proposed using $n$, $1 \times 1$; and $n$, $3 \times 3$ filters in the expand phase, where $n$ is $4\times$ larger than number of filters used in the squeeze phase. The entire SqueezeNet architecture is constructed by stacking conventional convolution layers, max-pooling, fire modules, and ends with an average pooling layer. The model has no fully connected layers. For more details about the number of fire modules for each stage, their order, and number of squeeze and expand filters for the different stages, see~\cite{iandola2016squeezenet}. 

\subsection{Inception} 
\label{inception} 
The Inception network is a deep convolutional architecture introduced as GoogLeNet (Inception V1) in 2014 by Szegedy et al.~\cite{googlenet2015}. The architecture has been refined in various ways such as adding batch normalization layers to accelerate training (Inception V2~\cite{ioffe2015batch}), and factorizing convolutions with larger spatial filters for computational efficiency (Inception V3~\cite{inceptionnet2016}). We adopt the Inception V3 model due to its outstanding performance in image recognition and object localization. 

\begin{figure}[tp] 
	\centering 
	{\includegraphics[width=7 cm, height=5 cm]{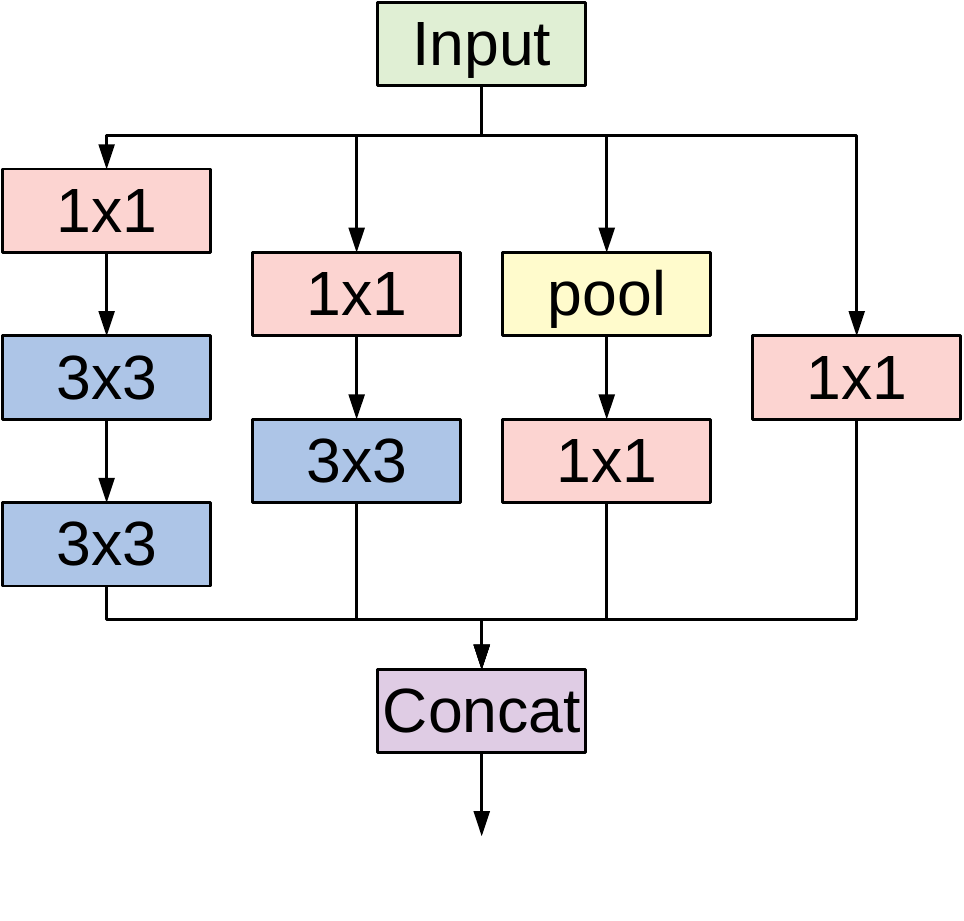}} 
	\caption{A variant of the Inception module used in InceptionV3 architecture.} 
	\label{inception_module} 
\end{figure}

The fundamental building block for all Inception-style networks is the Inception module of which several forms exist. Figure~\ref{inception_module} shows one variant of the Inception module that is used in the Inception V3 model. The module accepts an input and then branches into four different paths each performing a specific set of operations. The input passes through convolutional layers with different kernel sizes ($1\times 1$ and $3 \times 3$) as well as a pooling operation. Applying different kernel sizes allows the module to capture complex patterns at different scales. The outputs of all branches are concatenated channel-wise. 

The overall architecture of the Inception V3 network is composed of conventional $3 \times3$ convolutional layers at the early stages of the network, where some of these layers are followed by max-pooling operations. Subsequently a stack of various Inception modules is applied. These modules have different designs with respect to the number of applied filters, filter sizes, depth of the module after symmetric or asymmetric factorization of larger convolutions, and when to expand the filter bank outputs. The last Inception module is followed by an average-pooling operation and a fully connected layer. 

\subsection{ResNet} 
\label{ResNet} 
Deep Residual Networks (ResNet) proposed by He et al. in \cite{he2016deep}, represent a family of extremely deep CNN architectures that won the 2015 Large Scale Visual Recognition Challenge (ILSVRC-2015) for image recognition, object detection and localization~\cite{russakovsky2015imagenet}. The winning network is composed of 152 layers, which confirms the beneficial impact of network depth on visual representations. However, two major problems are encountered when training networks of increasing depth; vanishing gradients and performance degradation. The authors addressed the problems by adding skip connections to prevent information loss as the network gets deeper. 

\begin{figure}[tp] 
	\centering 
	\begin{subfigure}[] 
		{\includegraphics[width=3 cm, height=4 cm]{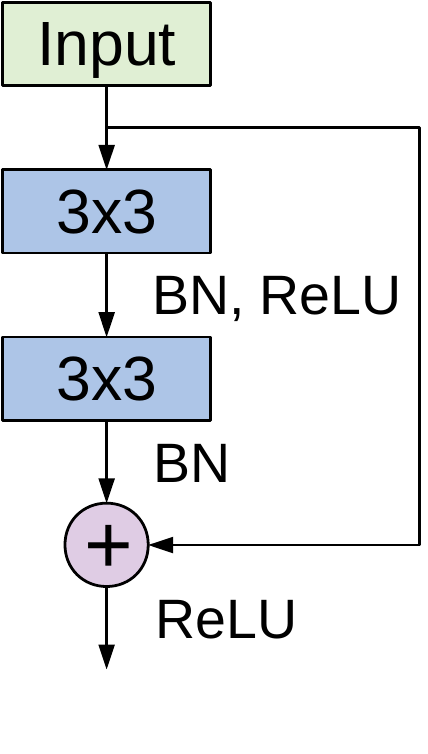}} 
	\end{subfigure} 
	~~~~~~~~~~~~~~~~~~~~ 
	\begin{subfigure}[] 
		{\includegraphics[width=3 cm, height=4 cm]{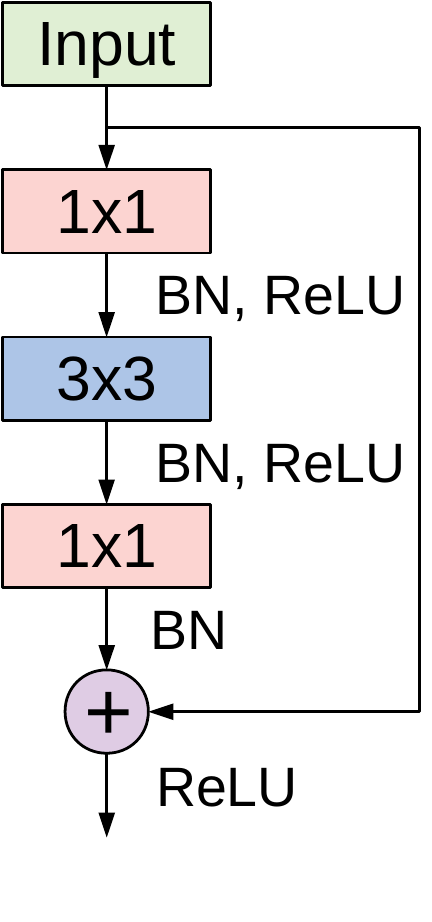}} 
	\end{subfigure} 
	\caption{The basic building block residual module employed in ResNet18 (a), and the bottleneck residual module used in ResNet50 and ResNet101 (b), both as introduced in~\cite{he2016deep}.} 
	\label{resnet_module} 
\end{figure}

The cornerstone for constructing deep residual networks is the residual module of which two variants are depicted in Figure~\ref{resnet_module}. The left path of the residual module in Figure~\ref{resnet_module} (a) is composed of two convolutional layers, which apply $3 \times 3$ kernels and preserve the spatial dimensions. Batch normalization and ReLU activation are also applied. The right path is the skip connection where the input is added to the output of the left path. This variant is used in the ResNet18 model. Another variant of the residual module named the bottleneck residual module is depicted in Figure~\ref{resnet_module} (b), in which the input signal also passes through two branches. However, the left path performs a series of convolutions using $1\times 1$ and $3 \times 3$ kernel sizes, along with batch normalization and ReLU activation. The right path is the skip connection, which connects the module's input to an addition operation with the output of the left path. This variant is utilized in ResNet50 and ResNet101 models. 

A deep residual network is constructed by stacking multiple residual modules on top of each other along with other conventional convolution and pooling layers. For our experiments we adopt three variants of ResNet, the ResNet18, ResNet50 and ResNet101 models. The full configurations and overall structure about each model are given in~\cite{he2016deep}. 

\subsection{ResNeXt} 
\label{resnext} 
\begin{figure}[!h] 
	\centering 
	{\includegraphics[width=0.9 \linewidth, height=6 cm]{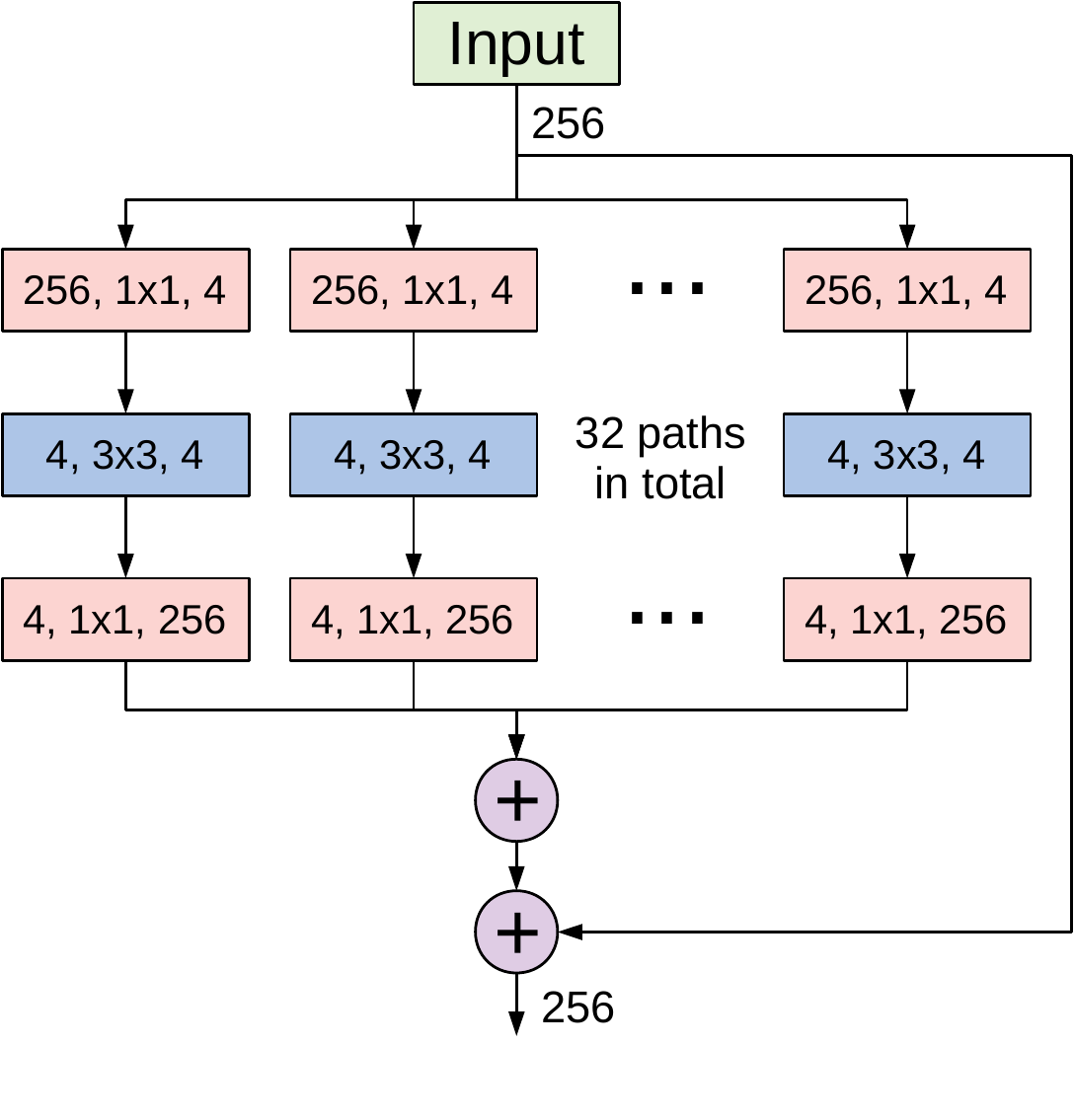}} 
	\caption{A ResNeXt building block with cardinality of 32~\cite{xie2017aggregated}.} 
	\label{resnext_module} 
\end{figure}

The ResNeXt architecture proposed in~\cite{xie2017aggregated} is a deep CNN model constructed by stacking residual building blocks of identical topology in a highly modularized fashion. Its simple design shares similarities with the ResNet architecture. ResNeXt also exploits the split-transform-merge strategy of the Inception module in an easy and extendable manner. The ResNeXt building block uses an identical set of transformations in every branch and hence allows the number of branches to be investigated as an independent hyperparameter. ResNeXt refers to the size of the set of transformations as the cardinality, which represents an important dimension for improving the network's representational power. Figure~\ref{resnext_module} depicts a ResNeXt building block with a cardinality of 32. Each branch applies the same set of transformations and their outputs are aggregated by summation. 

The entire network is constructed by stacking ResNeXt blocks along with other conventional convolution and pooling layers. For our experiments we implement two ResNeXt models, the 50-layer and the 101-layer networks. In a similar manner as their ResNet counterparts, ResNeXt models use RGB-inputs of size $224 \times 224$. However, we found an input size of $349 \times 253$, similar to the ResNet models, achieves the best performance on the considered datasets. 

\subsection{Xception} 
\label{xception} 
Xception is a deep CNN architecture proposed in~\cite{chollet2017xception}. It is inspired by the Inception architecture and utilizes the residual connections proposed in ResNet models~\cite{he2016deep}. However, it replaces the Inception modules with depthwise separable convolution layers. A depthwise separable convolution consists of a depthwise convolution (spatial convolution of $3 \times 3$, $5 \times5$, etc.) performed over each channel of an input to map the spatial correlations, followed by a pointwise convolution ($1 \times1$) to map the cross-channel correlations. 

The Xception architecture depends entirely on depthwise separable convolution layers with a strong assumption that spatial correlations and cross-channel correlations can be mapped separately. The network consists of 36 convolutional layers structured into 14 modules. All modules have residual connections except for the first and last modules. The reader is referred to \cite{chollet2017xception} for a complete description of the model specification. 

Due to its superior performance in vision tasks, we adopt the Xception model in our experiments. Even though the original model uses an RGB-input of size $299 \times 299$, we found that an input size of $327 \times 231$ obtains the best results. 

\subsection{ShuffleNet} 
\label{shufflenet} 
ShuffleNet is a very computationally-efficient CNN architecture that is mainly designed for mobile devices with constrained computational power~\cite{zhang2018shufflenet, ma2018shufflenet}. The architecture introduces two important operations to significantly reduce the computational cost while maintaining accuracy. The first operation is pointwise group convolutions, which can reduce the computational complexity of the $1 \times 1$ convolutions. The second operation consists of shuffling the channels, which assists the information flow across feature channels. 

\begin{figure}[tp] 
	\centering 
	{\includegraphics[width=4 cm, height=7 cm]{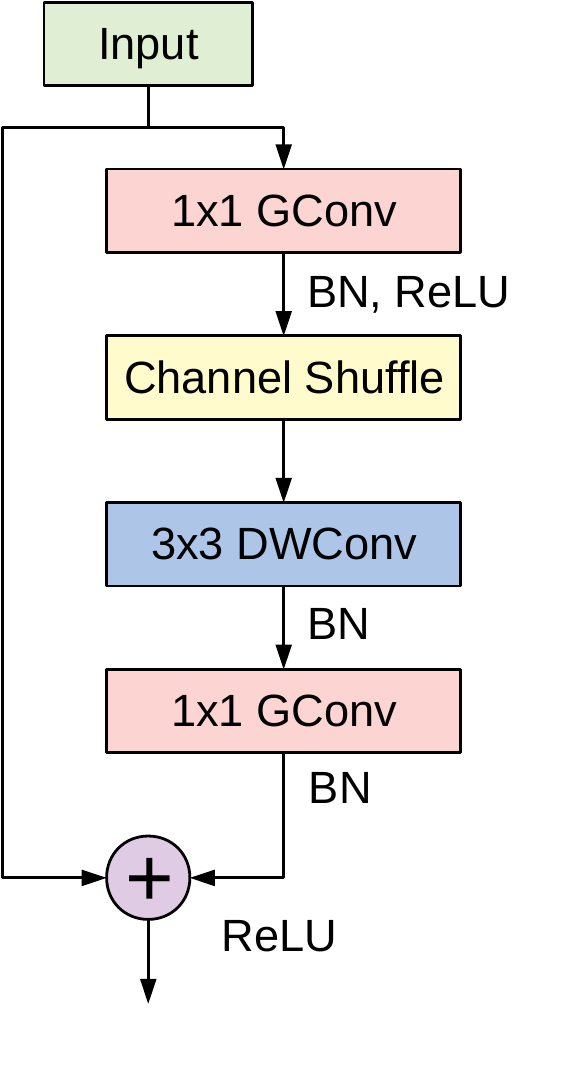}} 
	\caption{The building unit of the ShuffleNet architecture.} 
	\label{shufflenet_module} 
\end{figure}

The cornerstone of the ShuffleNet model is the ShuffleNet unit depicted in Figure~\ref{shufflenet_module}. It is a bottleneck residual module in which the $3 \times 3$ convolutional layer is replaced by a $3 \times 3$ depthwise separable convolution as in \cite{chollet2017xception}. Also, the first $1 \times 1$ convolutional layer is replaced by a pointwise group convolution followed by a channel shuffle operation. The second pointwise group convolutional layer is used to retrieve the channel dimension to match the left path of the unit. The overall ShuffleNet network is composed of a stack of these units grouped into three different stages along with other conventional convolution and pooling layers. 

In this study we adopt the recent variant of the ShuffleNet architecture. The original model uses an RGB-input of $224 \times 224$, however, we found that an input resolution of $321 \times 225$ works better for the considered datasets. 

\subsection{DenseNet} 
\label{densenet} 
Densely Connected Convolutional Networks (DenseNets) are a class of CNN architectures introduced in \cite{densenet2017} with several compelling characteristics. They alleviate the vanishing gradients problem, foster feature reuse, achieve high performance, consolidate feature propagation, and are computationally efficient. DenseNets modify the shortcut connections from ResNet by concatenating the output of the convolutions instead of summing them up. So, the input to the next layer will be the feature maps of all the preceding layers. 

\begin{figure}[!h] 
	\centering 
	{\includegraphics[width=0.85\linewidth]{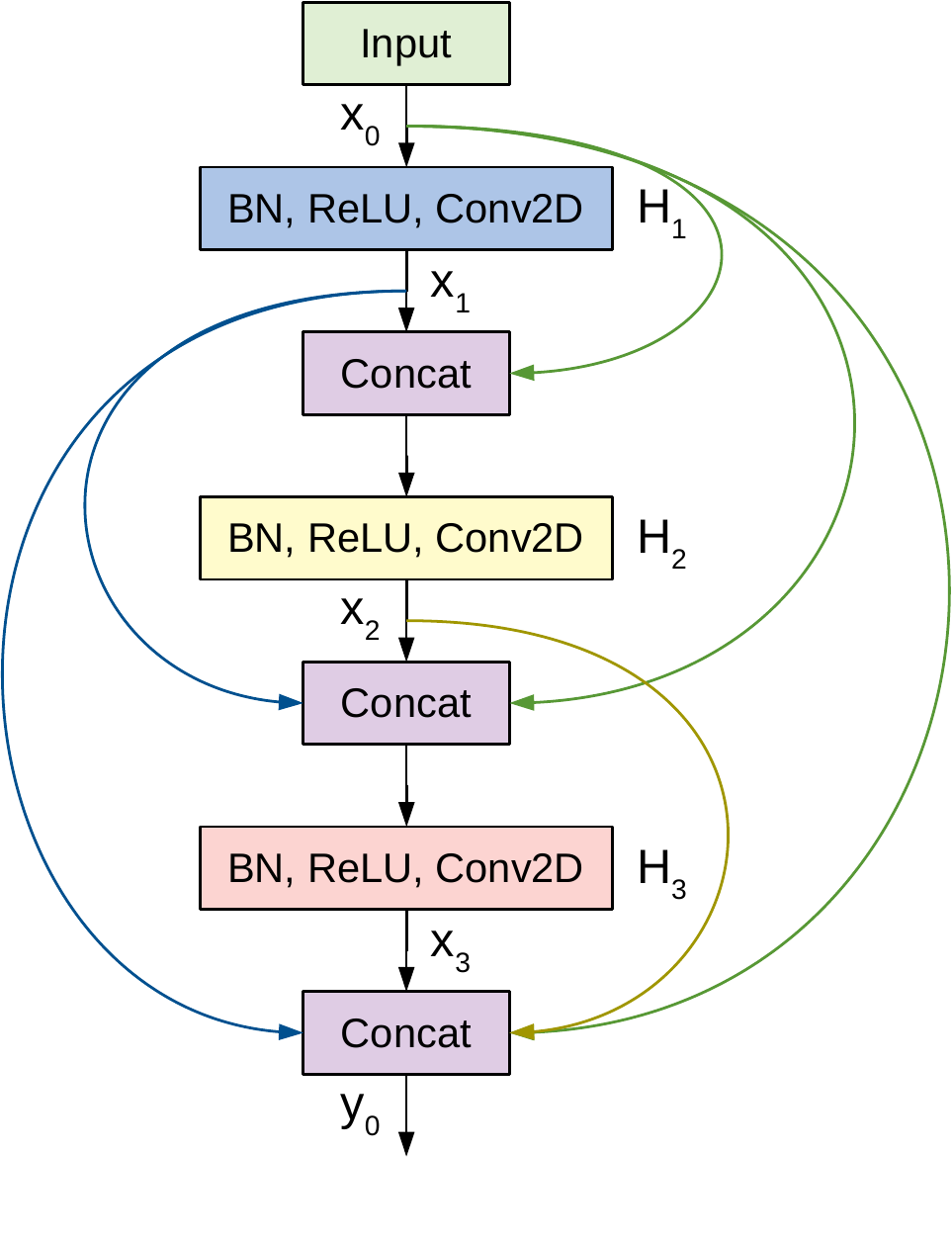}} 
	
	\caption{A 3-layer Dense block in DenseNet. The input to each layer is all the previous feature maps.} 
	\label{densenet_module} 
\end{figure}

Figure~\ref{densenet_module} shows a 3-layer Dense block where each layer performs a set of batch normalization (BN), ReLU activation and $3 \times 3$ Convolution operations. Previous feature maps are concatenated and presented as the input to a layer, which then generates $k$ feature maps. $k$ is a newly introduced hyper-parameter, denoted as the growth rate. Thus, if the input to layer $x_0$ is $k_{0}$, then the number of feature maps at the end of a 3-layer Dense block is $3\times k + k_0$. To prevent the number of feature maps from increasing too rapidly, DenseNet introduces a bottleneck layer with $1 \times 1$ convolution and  $4\times k$ filters. To tackle the difference in the feature map sizes when transitioning from a large feature map to a smaller one, DenseNet applies a transition layer made of $1 \times 1$ convolution and average pooling. 

A deep DenseNet is constructed by stacking multiple Dense blocks with transition layers. Conventional convolution and pooling layers are used at the beginning of the network. Eventually the output is pooled by Global average pooling, flattened and passed to a softmax classifier. For our study we experiment with three variants of DenseNet, the 121-layer, 169-layer and 201-layer architectures. The original models use an RGB-input of $224 \times 224$, however, we found that an input size of $349 \times 253$ achieves better results for images from the used datasets. 

Table~\ref{characteristics} summarizes the important characteristics of the adopted deep CNN models. This includes the square-sized input for each network, our proposed custom-sized input, trainable parameters in millions, number of layers and the model size in megabytes.
\begin{table*}[tp]
	\caption{Characteristics of the deep CNN architectures considered for this work.}
	\label{characteristics}      
	\centering
	\resizebox{\linewidth}{!}
	{
		\begin{tabular}{c|ccccc}	
			\hline
			\multirow{2}{*}{\textbf{Model}} & \multicolumn{5}{c}{\textbf{Model characteristics}}   \\
			\cline{2-6}	
			& \textbf{Default input size} & \textbf{Custom input size} & \textbf{Layers} &  \textbf{Parameters (M)} &\textbf{Model size (MB)} \\
			\hline
			SqueezeNet & $ 227 \times 227 $ & $ 335 \times 255$  & 18 & 0.73 & 3.0 \\
			ShuffleNet & $ 224\times 224$ & $ 321 \times 225 $ & 51 & 0.34 & 1.5\\
			ResNet18 & $ 224\times 224$  &  $ 349 \times 253 $ & 18 &  11.17 & 44.8\\
			ResNet50 & $ 224\times 224$ & $ 349 \times 253 $ & 50 & 23.51 & 94.3 \\
			ResNet101 & $ 224\times 224$ &  $ 349 \times 253 $ & 101 & 42.50 & 170.6 \\
			ResNeXt50 & $ 224\times 224$ & $ 349 \times 253 $ & 50 & 22.98 &  92.3\\
			ResNeXt101 & $ 224\times 224$ &  $ 349 \times 253 $ & 101 & 86.74 & 347.9\\
			InceptionV3 & $299\times 299$ & $ 331 \times 267 $  & 48 & 21.79 & 87.4\\
			Xception & $ 299\times 299$ & $ 327\times 231 $  & 37 & 20.81 & 83.5\\
			DenseNet121 & $ 224\times 224$ &$  349\times 253 $ & 121 & 6.95 & 28.3 \\
			DenseNet169 & $ 224\times 224$ & $  349\times 253$  & 169 & 12.48 &  50.8\\
			DenseNet201 & $ 224\times 224$ &  $ 349 \times 253$ & 201 &  18.09 &  73.6\\
			\hline		
		\end{tabular}
	}	
\end{table*}

\section{Transfer Learning}
\label{methodology}
Transfer learning is an effective representation learning approach in which the networks trained on abundant amount of images (millions) are used to initialize the networks for tasks for which data is scarce (a few hundreds or thousands of images). In the context of deep learning there are two common strategies to apply transfer learning from pretrained networks: feature extraction and fine-tuning \cite{huh2016makes, alshazly2019ensembles}. In the first strategy only the weights of some newly added layers are optimized during training, while in the second strategy all the weights are optimized for the new task. Here, we consider fine-tuning as a more effective strategy that outperforms feature extraction and achieves better performance. As our pretrained networks explicitly require an RGB-input, we assign identical values to the R, G and B channels. Since the CT images in the two datasets have varying spatial sizes, the images need to be scaled to match the target input size. One strategy to unify images with different aspect ratios involves stretching or excessive cropping. We opted for a different, less violating procedure and embed the image into a fixed-sized canvas. The aspect ratio of the original image is not altered and padding is applied to match the target shape.
\vspace{-0.5 cm}
\section{Experiments and Results}
\label{experiments}
This section presents our experimental setup and extensive experiments to show the efficacy of our fine-tuned networks. First, we describe the CT image datasets. Second, we state the experimental settings and performance evaluation metrics. Third, we discuss the obtained results of different models on each dataset. Finally, we apply two visualization methods to facilitate interpretation of the results and to localize the COVID-19 associated regions. 
\vspace{-0.5 cm}
\subsection{Datasets}
\label{datasets}

\begin{figure*}[]
\centering
\begin{subfigure}
	{\includegraphics[width=2 cm,height=2 cm]{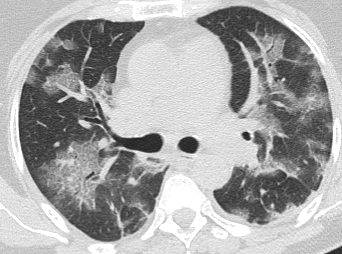}}
\end{subfigure}	
\begin{subfigure}
	{\includegraphics[width=2 cm,height=2 cm]{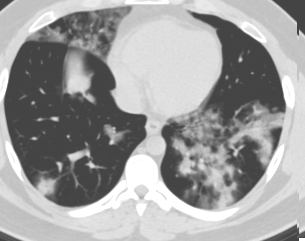}}
\end{subfigure}		
\begin{subfigure}
	{\includegraphics[width=2 cm,height=2 cm]{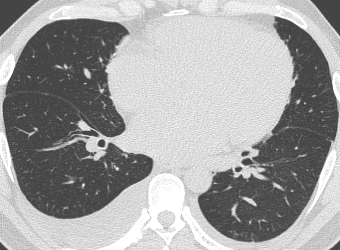}}
\end{subfigure}
\begin{subfigure}
	{\includegraphics[width=2 cm,height=2 cm]{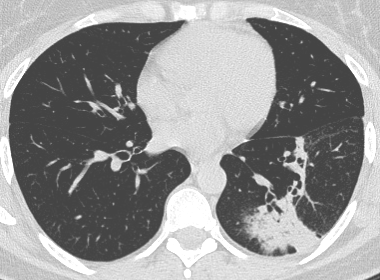}}
\end{subfigure}	
\begin{subfigure}
	{\includegraphics[width=2 cm,height=2 cm]{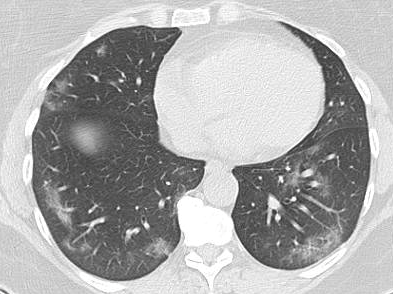}}
\end{subfigure}\\		
\begin{subfigure}
	{\includegraphics[width=2 cm,height=2 cm]{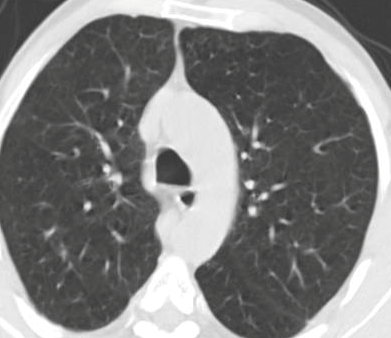}}
\end{subfigure}	
\begin{subfigure}
	{\includegraphics[width=2 cm,height=2 cm]{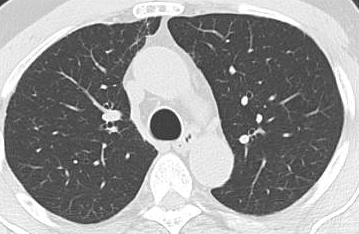}}
\end{subfigure}		
\begin{subfigure}
	{\includegraphics[width=2 cm,height=2 cm]{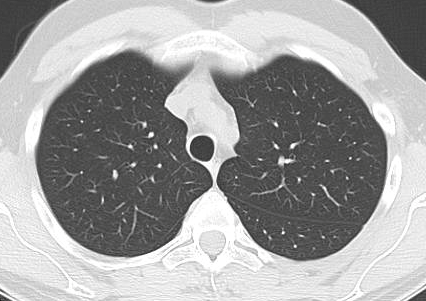}}
\end{subfigure}
\begin{subfigure}
	{\includegraphics[width=2 cm,height=2 cm]{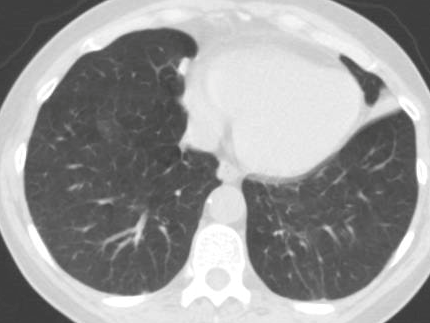}}
\end{subfigure}	
\begin{subfigure}
	{\includegraphics[width=2 cm,height=2 cm]{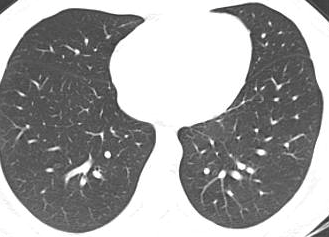}}
\end{subfigure}		
	\caption{Examples of chest CT scans from the SARS-CoV-2 dataset. The first row represents scans diagnosed with COVID-19, whereas the second row represents Non-COVID-19 but other lung diseases.}
	\label{sars-cov-2-ct_images} 
\end{figure*}

\textbf{SARS-CoV-2 CT Scan dataset}~\cite{soares2020sars}: The dataset was collected from hospitals of Sao Paulo, Brazil, with a total of 2482 CT scans acquired from 120 patients of both genders. It is composed of 1252 scans for patients infected with SARS-CoV-2 and 1230 scans for patients infected with other lung diseases. The CT scans have varying spatial sizes between $119 \times 104$ and $416 \times 512$, and are available in PNG format. CT scans from this dataset are shown in Figure~\ref{sars-cov-2-ct_images}.

\begin{figure*}[ht]
\centering
\begin{subfigure}
	{\includegraphics[width=2 cm,height=2 cm]{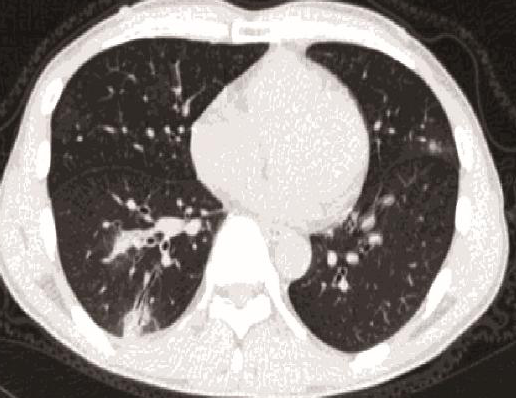}}
\end{subfigure}	
\begin{subfigure}
	{\includegraphics[width=2 cm,height=2 cm]{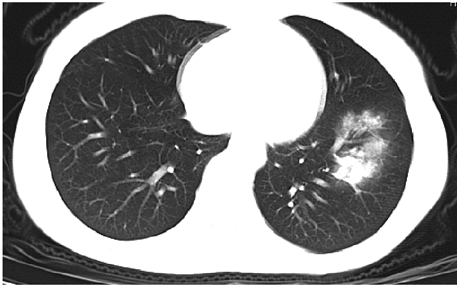}}
\end{subfigure}		
\begin{subfigure}
	{\includegraphics[width=2 cm,height=2 cm]{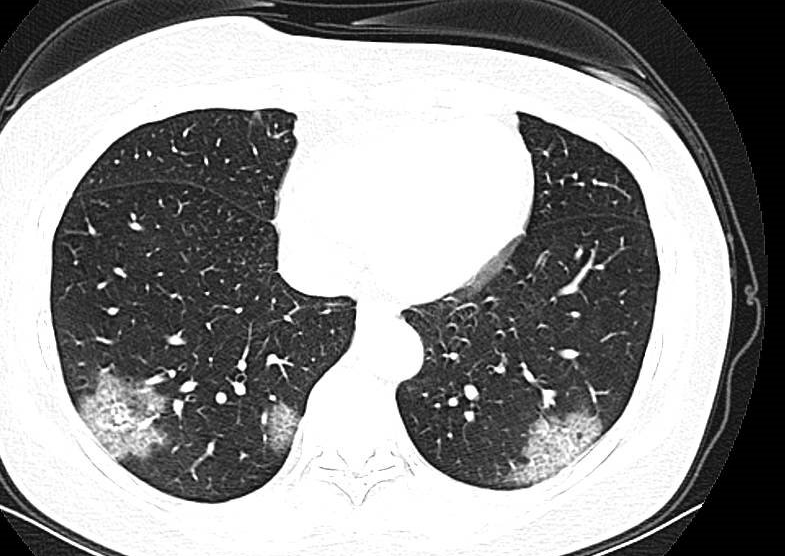}}
\end{subfigure}
\begin{subfigure}
	{\includegraphics[width=2 cm,height=2 cm]{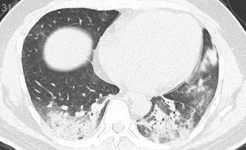}}
\end{subfigure}	
\begin{subfigure}
	{\includegraphics[width=2 cm,height=2 cm]{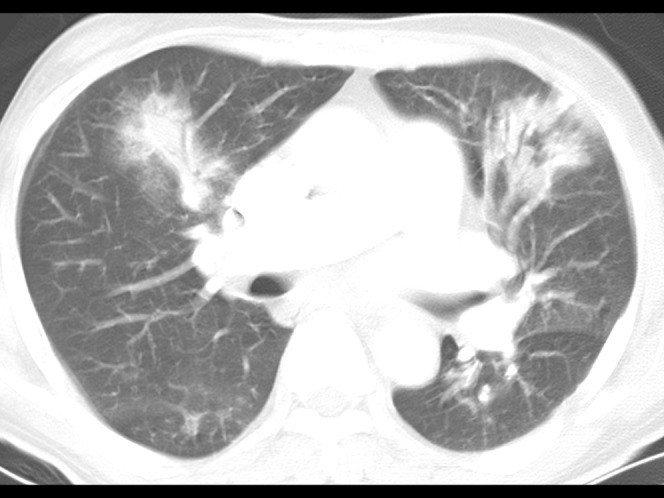}}
\end{subfigure}	\\	
\begin{subfigure}
	{\includegraphics[width=2 cm,height=2 cm]{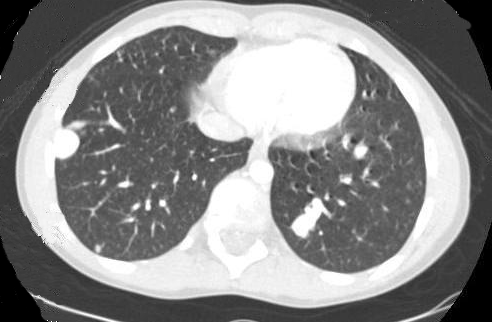}}
\end{subfigure}	
\begin{subfigure}
	{\includegraphics[width=2 cm,height=2 cm]{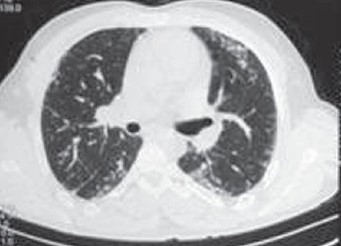}}
\end{subfigure}		
\begin{subfigure}
	{\includegraphics[width=2 cm,height=2 cm]{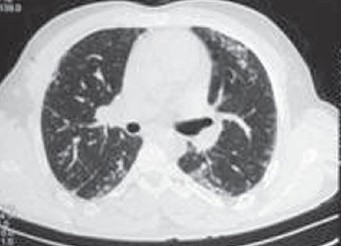}}
\end{subfigure}
\begin{subfigure}
	{\includegraphics[width=2 cm,height=2 cm]{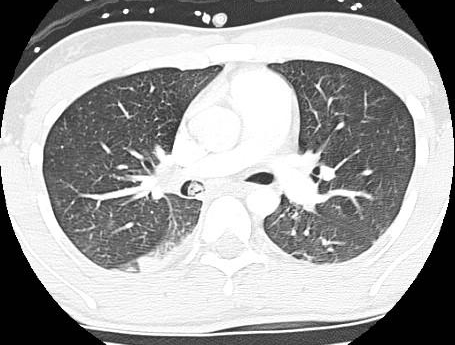}}
\end{subfigure}	
\begin{subfigure}
	{\includegraphics[width=2 cm,height=2 cm]{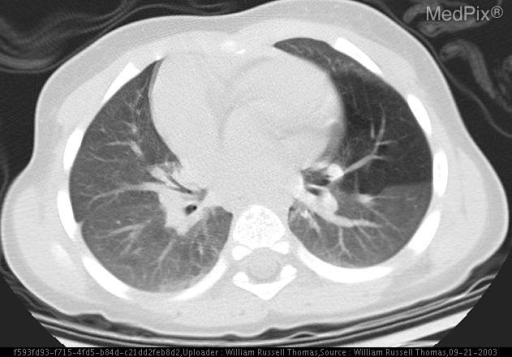}}
\end{subfigure}		
	\caption{Examples of chest CT images from the COVID19-CT dataset. The first row represents images diagnosed with COVID-19, whereas the second row represents Non-COVID-19 cases.}
	\label{covid19-ct_images} 
\end{figure*}

\textbf{COVID19-CT dataset}~\cite{he2020sample}: The dataset consists of a total of 746 CT images. There are 349 CT images of patients with COVID-19 and 397 CT images showing Non-COVID-19, but other pulmonary diseases. The positive CT images were collected from preprints about COVID-19 on medRxiv and bioRxiv, and they feature various manifestations of COVID-19. Since the CT images were taken from different sources, they have varying sizes between $124 \times 153$ and $1485 \times 1853$. Figure~\ref{covid19-ct_images} shows example CT images from the COVID19-CT dataset.

\subsection{Experimental Settings}
\label{expiremental_settings}
To assess the performance of our models we perform five-fold cross-validation. The final performance of the models is computed by averaging the obtained values from the five networks on their test fold respectively. 

Data augmentation methods are implemented to effectively increase the amount of training samples for improved generalization. Affine transformations like rotation and shearing turned out to have a worsening effect on performance, so we excluded this type of augmentations. More augmentation steps include cropping, adding blur with a probability of $25\%$, adding a random amount of Gaussian noise, changes in brightness and contrast and random horizontal flipping. Finally, the images are normalized according to the ImageNet dataset.

We follow a set of optimization configurations for all deep networks. The networks are optimized by applying the LAMB optimizer \cite{you2020large} on a binary cross-entropy loss. The initial learning rate is set to 0.0003 and is scheduled to decrease according to the following steps: epoch 50: 0.0001, epoch 70: 0.00003, epoch 80: 0.00001, epoch 90: 0.000003. We use a batch size of 32 and we apply a high weight decay of 1 for regularization. The networks are implemented using the PyTorch framework and are trained for 100 epochs on a PC with Intel(R) Core(TM) i7-3770 CPU, 8 MB RAM and Nvidia GTX 1080 GPU. 

\subsection{Evaluation Metrics}
\label{metrics}
We consider different performance evaluation metrics for evaluating our models. For each model we count the number of predicted cases as True Positives (TP), True Negatives (TN), False Positives (FP) and False Negatives (FN). Then, we compute the following metrics.

\begin{equation}
Accuracy=\frac{\text{TP} + \text{TN}}{\text{TP} + \text{TN} + \text{FP} + \text{FN}}
\end{equation}

\begin{equation}
Precision=\frac{\text{TP}}{\text{TP} + \text{FP}}
\end{equation}

\begin{equation}
Recall\: (sensitivity)=\frac{\text{TP}}{\text{TP} + \text{FN}}
\end{equation}

\begin{equation}
Specificity=\frac{\text{TN}}{\text{TN} + \text{FP}}
\end{equation}

\begin{equation}
F1{-}score=2\times \frac{\text{Precision} \times \text{Recall}}{\text{Precision} + \text{Recall}}
\end{equation}

\vspace{-0.5 cm}
\subsection{Results and Discussion}
\label{results_analysis}
Here, we present and discuss the obtained results for detecting COVID-19 on the considered CT image datasets with different deep networks. We report the quantitative results along with the confusion matrices for every single architecture of the adopted networks. 

Table~\ref{results} summarizes the average values of the evaluation metrics achieved by different deep networks on the two CT image datasets. All values are given in percentages and the best results are written in bold. We also compare with the previously obtained results from the literature when applicable. Generally, we observe some performance differences between the obtained results on the SARS-CoV-2 CT and the COVID19-CT datasets. Also, we observe the superiority of our model compared with similar models from recently published works, which indicates the effectiveness of our optimization and learning strategy. 

\begin{table*}[tp]
	\caption{Performance comparison of different deep models for detecting COVID-19 using various evaluation metrics. The results are given in the form of mean and standard deviation scores. For a direct comparison, the results from recently published works are included when applicable.}
	\label{results}      
	\centering
	\resizebox{\linewidth}{!}
	{
		\begin{tabular}{c|c|ccccc}	
			\hline
		    \multirow{2}{*}{\textbf{Dataset}} & \multirow{2}{*}{\textbf{Model}} & \multicolumn{5}{c}{\textbf{Evaluation Metrics}}   \\
			\cline{3-7}	
			& & \textbf{Accuracy} & \textbf{Precision} & \textbf{Recall} &  \textbf{Specificity} &\textbf{F1-score} \\
			\hline
			 \multirow{15}{*}{SARS-CoV-2 CT}& SqueezeNet & $95.1 \pm 1.3$ & $94.2\pm 2,0$ & $96.2 \pm 1.4$  & $94.0 \pm 2.2$ & $95.2 \pm 1.2$ \\
			 & ShuffleNet & $97.5 \pm 0.8$ & $96.1 \pm 1.4$ & $99.0\pm 0.2$ & $95.9 \pm 1.5$ & $97.5 \pm 0.8$ \\
		     & ResNet18 & $98.3 \pm 0.8$  & $97.2 \pm 1.2$ & $99.6 \pm 0.3$ & $97.1 \pm 1.4$ & $98.4 \pm 0.7$ \\
			 & ResNet50 & $99.2 \pm 0.3$ & $99.1 \pm 0.5$ & $99.4 \pm 0.5$ & $99.1 \pm 0.5$ & $99.2 \pm 0.3$\\
			 & ResNet101 & $\mathbf{99.4 \pm 0.4}$ & $\mathbf{99.6 \pm 0.3}$ & $99.1 \pm 0.6$ & $\mathbf{99.6 \pm 0.3}$& $\mathbf{99.4 \pm 0.4}$ \\
			 & ResNeXt50 & $99.1 \pm 0.5$ & $99.0 \pm 0.5$ & $99.3 \pm 0.5$  & $98.9 \pm 0.6$ & $99.1 \pm 0.5$ \\
			 & ResNeXt101 & $99.2 \pm 0.3$ & $99.2 \pm 0.4$ & $99.3 \pm 0.5$ & $99.2 \pm 0.4$& $99.2 \pm 0.3$\\ 
			 & InceptionV3 & $99.1 \pm 0.5$ & $98.5 \pm 0.8$ & $\mathbf{99.8 \pm 0.3}$ & $98.5 \pm 0.8$ & $99.1 \pm 0.5$ \\
			 & Xception & $98.8 \pm 0.6$ & $99.0 \pm 1.0$ & $98.6 \pm 1.1$  & $98.9 \pm 1.1$& $98.8 \pm 0.6$\\
			 & DenseNet121 & $99.3 \pm 0.3$ & $99.4 \pm 0.2$ & $99.2 \pm 0.5$  & $99.4 \pm 0.2$ & $99.3 \pm 0.3$ \\
			 & DenseNet169 & $99.3 \pm 0.5$ & $99.4 \pm 0.6 $ & $99.3 \pm 0.5$ & $99.3 \pm 0.7$ & $99.3\pm 0.4 $ \\
			 & DenseNet201 & $99.2 \pm 0.2$ & $99.0 \pm 0.4$ & $99.4 \pm 0.2$ & $98.9 \pm 0.4$ & $99.2 \pm 0.2$ \\ 
			 \cline{2-7}	
			 & xDNN~\cite{soares2020sars} & 97.3 & 99.1 & 95.5 & - & 97.3 \\
			 & DenseNet201~\cite{jaiswal2020classification} & 96.2 & 96.2 & 96.2 & 96.2 & 96.2 \\  
			 & Modified VGG19~\cite{panwar2020deep} & 95.0 & 95.3 & 94.0 & 94.7 & 94.3\\ 
			 & COVID CT-Net~\cite{yazdani2020covid} & - & - & $85.0 \pm 0.2$ & $96.2 \pm 0.1$ & $90.0 \pm 0.1$ \\
			 & Contrastive Learning~\cite{wang2020contrastive} & $90.8 \pm 0.9$ & $95.7 \pm 0.4$ & $85.8 \pm 1.1$ &-& $90.8 \pm 1.3$ \\	 
			\hline
			\multirow{14}{*}{COVID19-CT}& SqueezeNet & $87.3 \pm 3.2$& $86.3 \pm 6.1$& $86.5 \pm 2.3$ & $87.9 \pm 6.3$ & $86.5 \pm 3.0$ \\
			& ShuffleNet & $87.9 \pm 2.6$ & $84.5 \pm 2.5 $ & $90.8 \pm 3.9$ & $85.4 \pm 2.7$ & $87.6 \pm 2.8$ \\
			& ResNet18 & $90.3 \pm 2.5$ & $87.1 \pm 4.1$ & $93.1 \pm 2.5$ & $87.9 \pm 4.9$& $90.1 \pm 2.3$ \\
			& ResNet50 & $90.8 \pm 1.9$ & $90.2 \pm 5.0$& $90.0 \pm 3.6$ & $91.4 \pm 5.0$ & $90.1 \pm 1.9$ \\ 
			& ResNet101 & $89.8\pm 2.5$ & $88.0 \pm 3.7$ & $90.5 \pm 1.9$ & $89.2 \pm 3.8$&  $89.3 \pm 2.4$ \\
			& ResNeXt50 & $90.6 \pm 2.2$& $87.4 \pm 3.6$ & $93.4 \pm 3.4$ & $88.2 \pm 4.4$& $90.3 \pm 2.2$  \\
			& ResNeXt101 & $90.9 \pm 1.8$& $88.1 \pm 3.5$ & $93.1 \pm 2.9$ & $88.9 \pm 4.0$& $90.6 \pm 1.8$  \\ 
			& InceptionV3 & $89.4 \pm 2.0$& $87.7 \pm 2.5$& $90.0 \pm 2.4$ & $88.9 \pm 2.4$& $88.8 \pm 2.2$\\
			& Xception & $88.5 \pm 2.6$& $87.3 \pm 2.7$ & $88.3 \pm 4.7$ & $88.7 \pm 2.9$& $87.7 \pm 2.9$ \\ 
			& DenseNet121 & $88.9 \pm 1.2$& $87.6 \pm 2.6$ &  $88.8 \pm 1.4$& $88.9 \pm 2.9$& $88.2 \pm 1.0$ \\
			& DenseNet169 & $91.2 \pm 1.4$ & $88.1 \pm 2.5$ & $93.7 \pm 1.2$ & $88.9 \pm 2.7$ & $90.8 \pm 1.4$ \\ 
			& DenseNet201 & $\mathbf{92.9 \pm 2.2}$& $\mathbf{91.3 \pm 2.2}$& $\mathbf{93.7 \pm 3.4}$& $\mathbf{92.2 \pm 2.2}$ & $\mathbf{92.5 \pm 2.4}$ \\ 
			\cline{2-7}	
		 	& DenseNet169~\cite{he2020sample} & 83.0 & - & - & - & 81.0 \\ 
		 	& Decision function~\cite{mishra2020identifying} & 88.3 & -& -&-&  86.7\\
		 	& ResNet101~\cite{saqib2020covid19} & 80.3 &78.2 &85.7&-&81.8\\ 
		 	& DenseNet121+SVM~\cite{jokandan2020uncertainty} & $85.9 \pm 5.9$ &-& $84.9 \pm 8.4$  & $86.8 \pm 6.3$ &-\\ 
		 	& DenseNet169~\cite{martinez2020classification} & $87.7 \pm 4.7$ & $90.2 \pm 6.0$ & $85.6 \pm 6.7$ &-& $87.8 \pm 5.0$ \\ 
		 	& Contrastive Learning~\cite{wang2020contrastive} & $78.6 \pm 1.5$ & $78.0 \pm 1.3$ &$79.7 \pm 1.4$&-& $78.8 \pm 1.4$\\
		 	\hline			
		\end{tabular}
	}	
\end{table*}

\begin{figure*}[!h]
	\centering
	\begin{subfigure}[SqueezeNet]
		{\includegraphics[width=5 cm,height=4 cm]{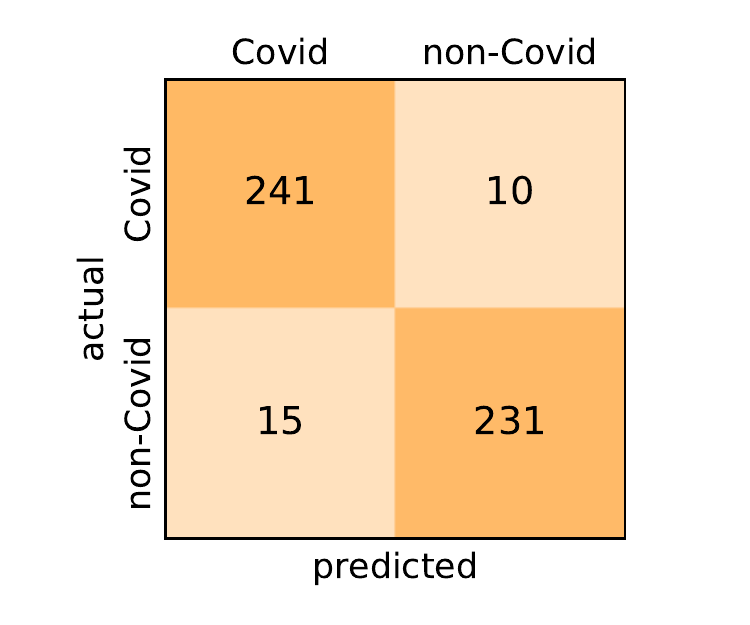}}
	\end{subfigure}	
	\begin{subfigure}[ShuffleNet]
		{\includegraphics[width=5 cm,height=4 cm]{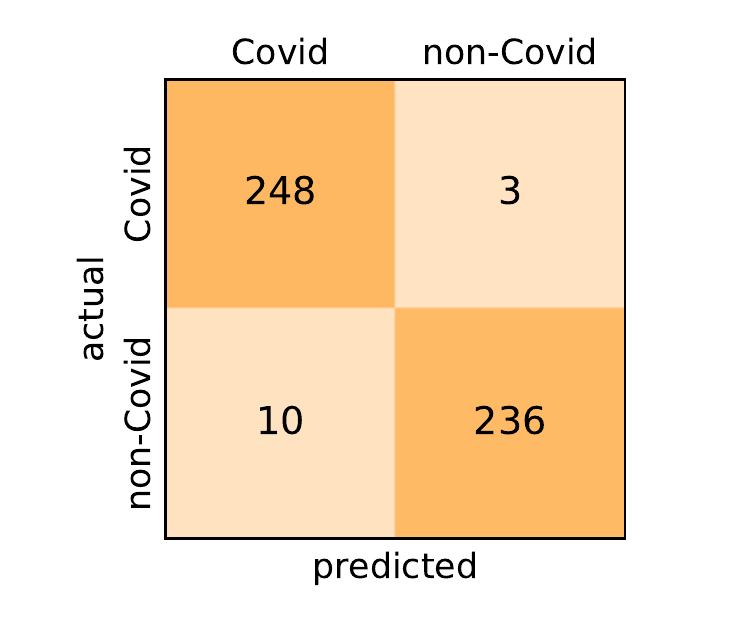}}
	\end{subfigure}		
	\begin{subfigure}[ResNet18]
		{\includegraphics[width=5 cm,height=4 cm]{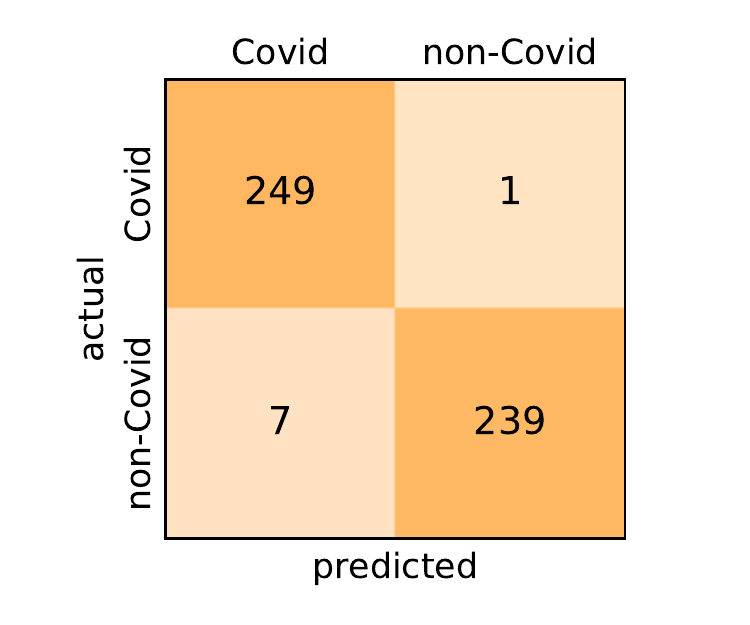}}
	\end{subfigure}
	\begin{subfigure}[ResNet50]
		{\includegraphics[width=5 cm,height=4 cm]{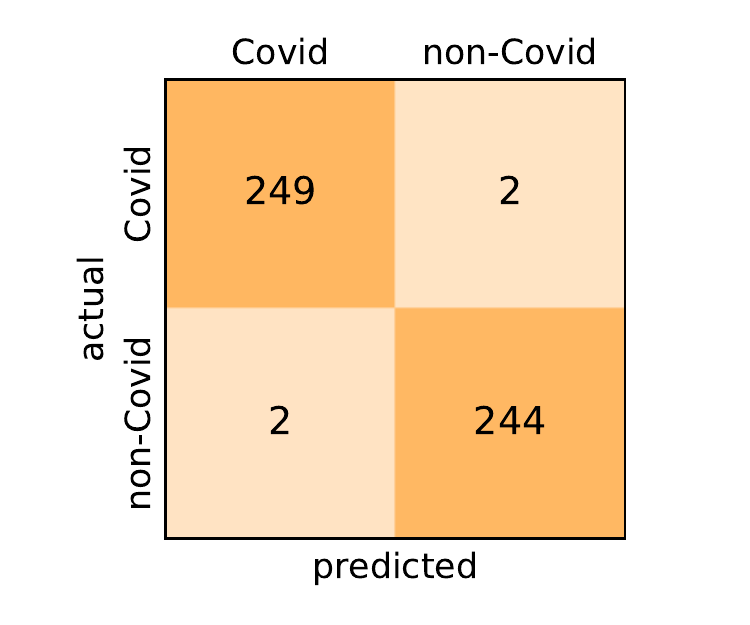}}
	\end{subfigure}	
	\begin{subfigure}[ResNet101]
		{\includegraphics[width=5 cm,height=4 cm]{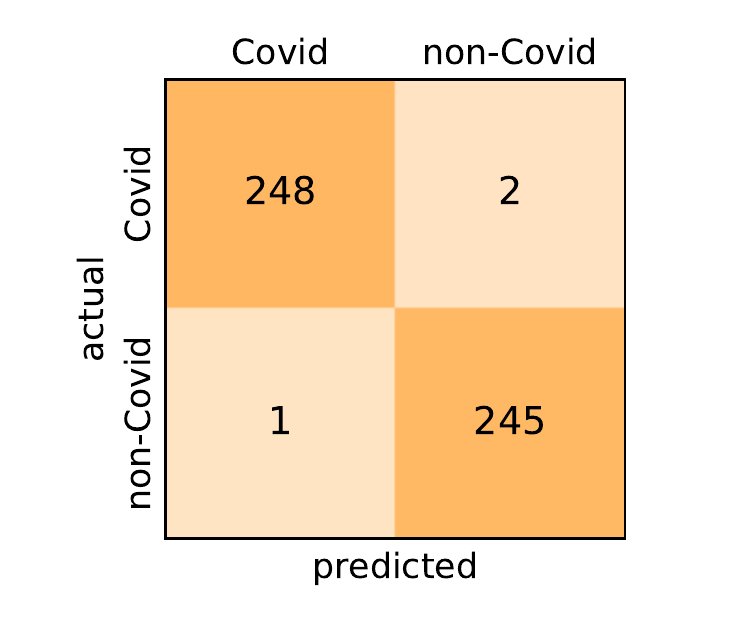}}
	\end{subfigure}		
	\begin{subfigure}[ResNeXt50]
		{\includegraphics[width=5 cm,height=4 cm]{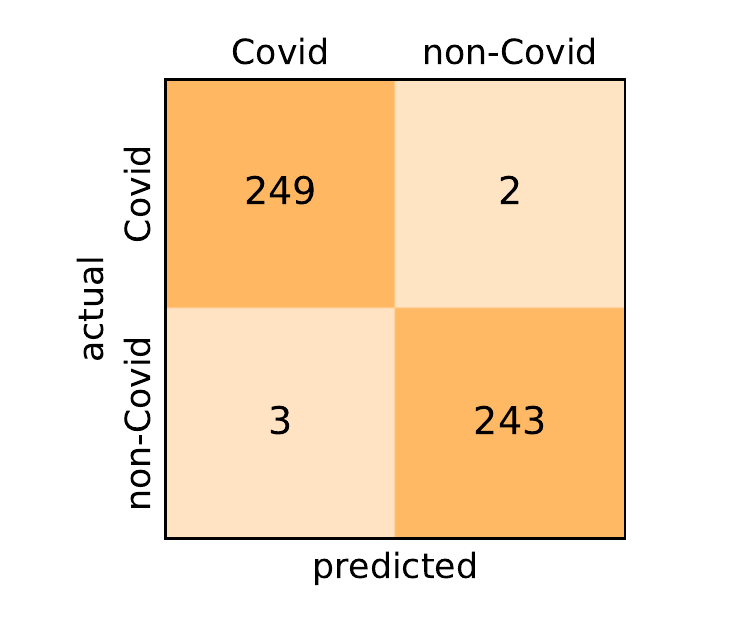}}
	\end{subfigure}	
	\begin{subfigure}[ResNeXt101]
		{\includegraphics[width=5 cm,height=4 cm]{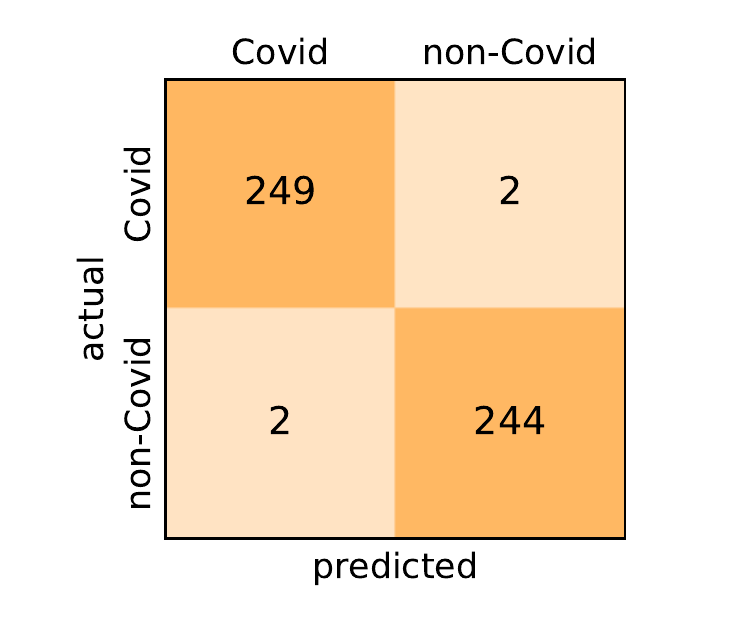}}
	\end{subfigure}		
	\begin{subfigure}[InceptionV3]
		{\includegraphics[width=5 cm,height=4 cm]{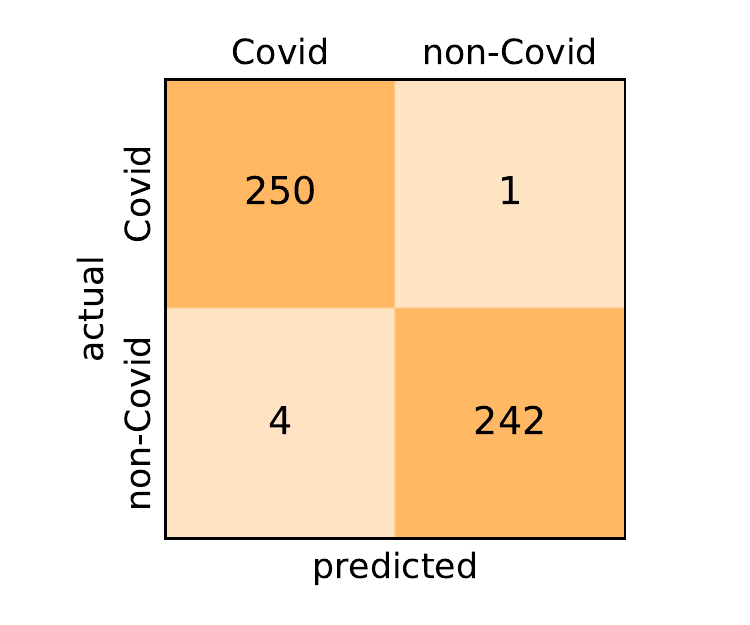}}
	\end{subfigure}
	\begin{subfigure}[Xception]
		{\includegraphics[width=5 cm,height=4 cm]{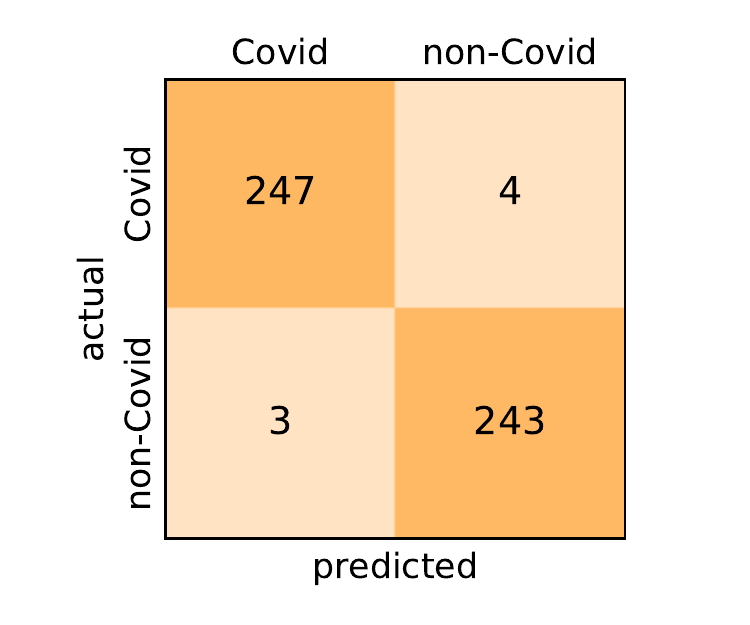}}
	\end{subfigure}	
	\begin{subfigure}[DenseNet121]
		{\includegraphics[width=5 cm,height=4 cm]{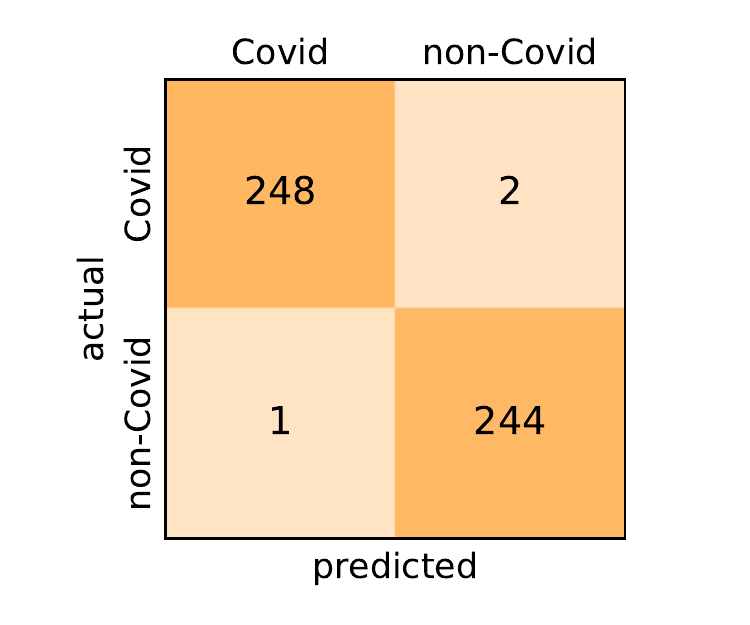}}
	\end{subfigure}	
	\begin{subfigure}[DenseNet169]
		{\includegraphics[width=5 cm,height=4 cm]{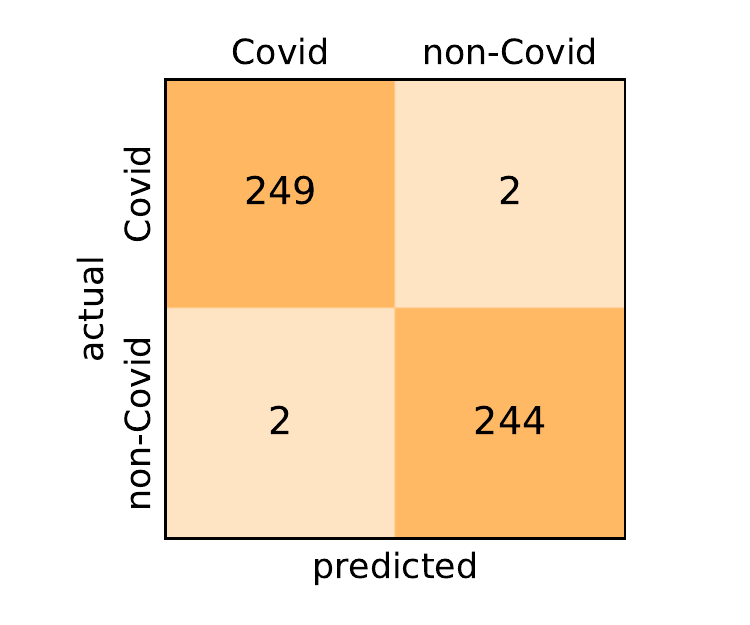}}
	\end{subfigure}	
	\begin{subfigure}[DenseNet201]
		{\includegraphics[width=5 cm,height=4 cm]{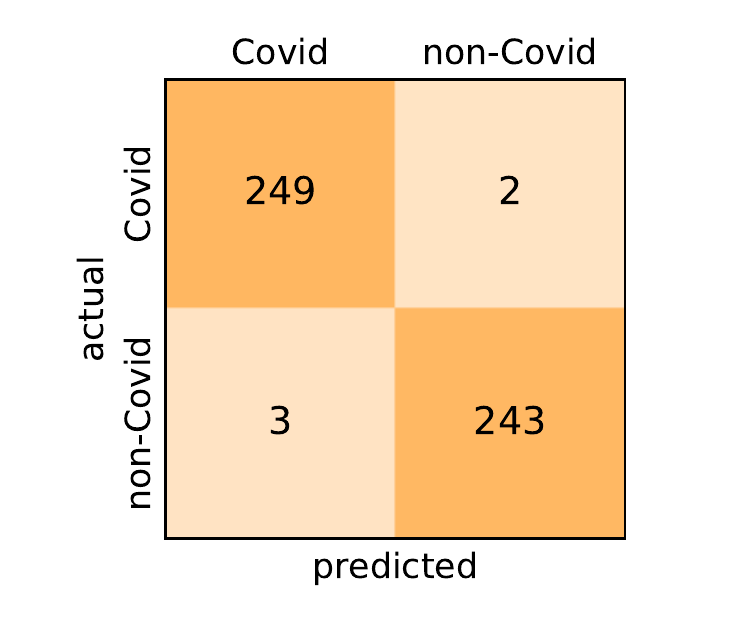}}
	\end{subfigure}	
	\caption{Confusion matrices for the different deep CNN models. These results are the average counts of the five models obtained by 5-fold cross-validation on the SARS-CoV-2 CT dataset.}
	\label{confusion_matrices} 
\end{figure*}

On the SARS-CoV-2 CT dataset, ResNet101 achieves the best overall performance with respect to almost all evaluation metrics, with an average accuracy and F1-score of $99.4\%$ and  $99.4\%$, respectively. The model also achieves an average sensitivity rate of $99.1\%$ indicating that, on average, only two COVID-19 images are falsely predicted as negatives. It is also powerful enough to correctly identify all Non-COVID-19 cases with only one false positive resulting a specificity rate of $99.6\%$. The highest sensitivity score of $99.8\%$ is achieved by the InceptionV3 model, where only one COVID-19 image is falsely predicted as negative on average. The SqueezeNet model obtains the lowest performance with respect to all evaluation metrics with a fairly acceptable average accuracy and sensitivity scores of $95.1\%$ and $96.2\%$, respectively. Also the ShuffleNet architecture obtains satisfactory performance with approximately $2\%$ improvements on average for all metrics compared with SqueezeNet. Although the results obtained by these models are inferior compared with the rest of models, but they are more efficient. This matches their main objective of reducing the computational costs rather than improving their visual recognition abilities. The rest of models achieve competitive performance and very promising results with slight performance differences. Comparing the different variants of ResNet and DenseNet, we can see that the deeper variants from each architecture yield a slightly better performance. The deeper ResNet101 and ResNeXt101 show a marginal gain in performance compared with their shallower counterparts. The details about class-wise results for each model are summarized in the confusion matrices in Figure~\ref{confusion_matrices}. 

It is worthy mentioning that on the SARS-CoV-2 CT dataset the inter-fold variations are minimal and usually below one percent, showing the robustness of our fine-tuning strategy. For some of the architectures like the DenseNet variants we observe a larger confidence interval than their actual differences in recognition performance. This means that the DenseNets and the deeper ResNet variants share a very similar performance and are almost indistinguishable from each other. Overall, the obtained results by our models are better than the recently published ones even when using the same network architectures. We attribute this to the better optimization and transferability of the learned features when applying our fine-tuning strategy.


On the COVID19-CT dataset, the overall performance with respect to all evaluation metrics is inferior to that on the SARS-CoV-2 dataset. This can be attributed to the cross-source heterogeneity of the CT images in the dataset. The Non-COVID-19 CT images were taken from different sources and show diverse findings which pose difficulty to distinguish between COVID-19 and other findings associated with lung diseases due to the potential overlap of visual manifestations (see Figure~\ref{covid19-ct_images}). Another reason is that, the CT images in the COVID19-CT dataset show strong variations in contrast, variable spatial resolution and other visual characteristics, which could affect the model’s ability to extract more discriminative and generalizable features.  

It is also worthy mentioning that for the COVID19-CT dataset the inter-fold variations grow substantially due to the small size of the dataset. During 5-fold cross-validation the training set consists of about 600 images only and the test fold has less than 200 images, which has to produce statistical fluctuations. Metrics considering the overall performance like the accuracy have less inter-fold variation. However, we observe stronger variations in metrics, that test the bias towards one of the classes like the specificity. The standard deviation of the specificity indicates that the different folds tend to encourage the model to focus more on COVID or more on Non-COVID cases. This phenomenon occurs even for stratified 5-fold cross-validation, where the distribution of classes in each fold represents the class distribution of the entire dataset, and it seems to originate from the small number of images only. 

\begin{figure*}[!h]
	\centering
	\begin{subfigure}[SqueezeNet]
		{\includegraphics[width=5 cm,height=4 cm]{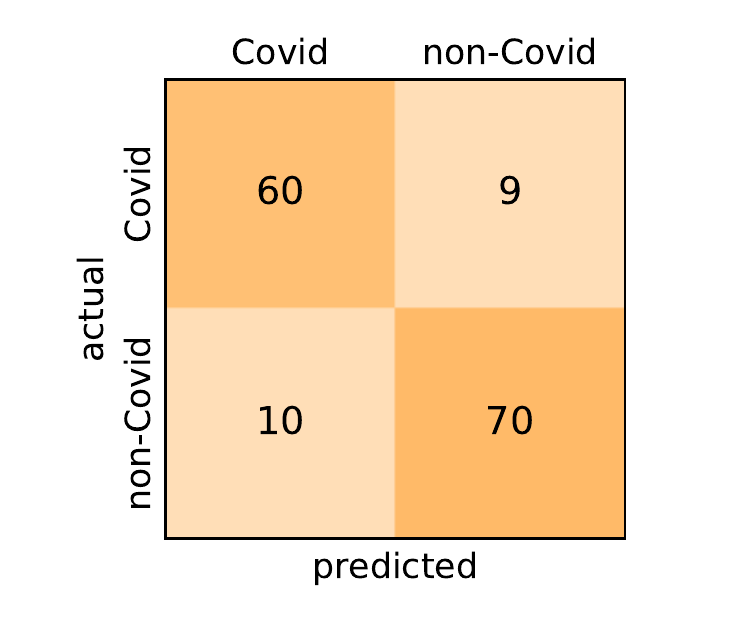}}
	\end{subfigure}	
	\begin{subfigure}[ShuffleNet]
		{\includegraphics[width=5 cm,height=4 cm]{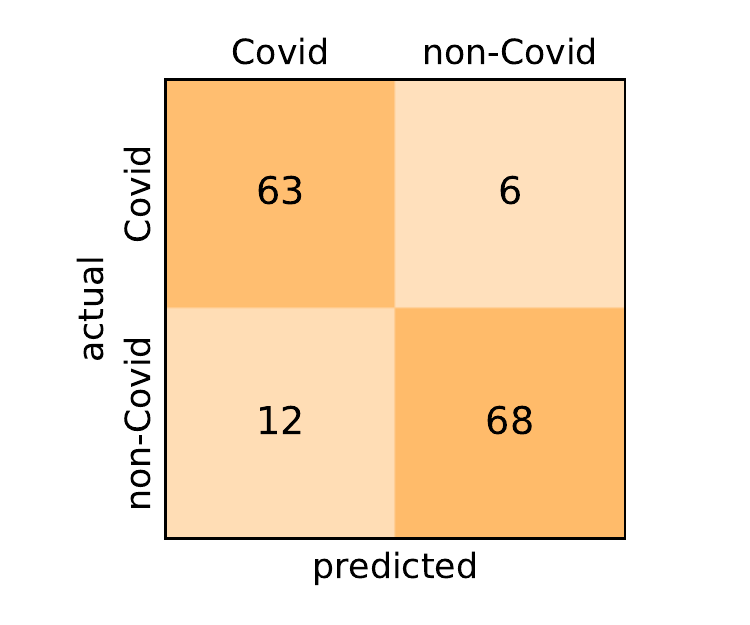}}
	\end{subfigure}		
	\begin{subfigure}[ResNet18]
		{\includegraphics[width=5 cm,height=4 cm]{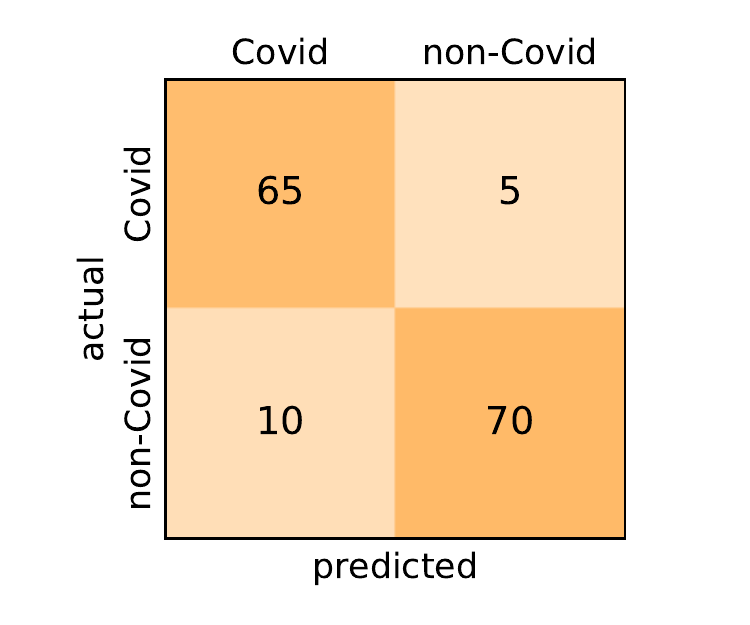}}
	\end{subfigure}
	\begin{subfigure}[ResNet50]
		{\includegraphics[width=5 cm,height=4 cm]{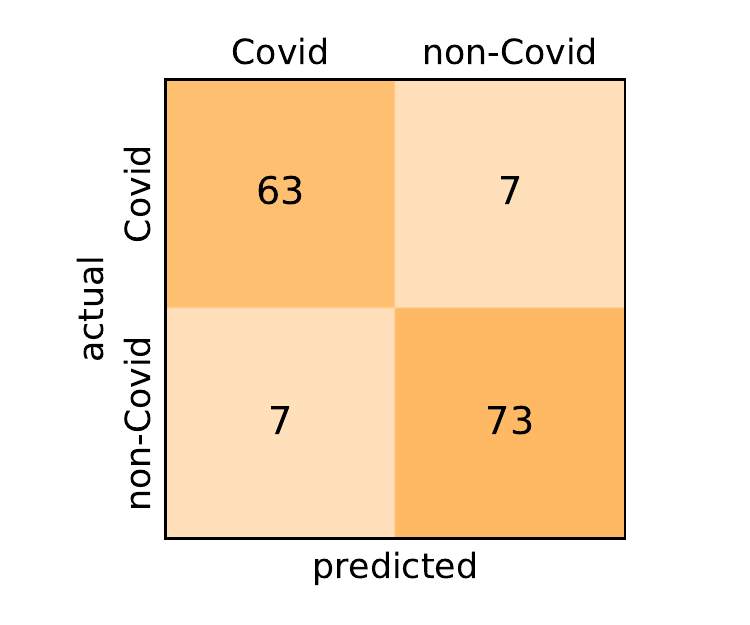}}
	\end{subfigure}	
	\begin{subfigure}[ResNet101]
		{\includegraphics[width=5 cm,height=4 cm]{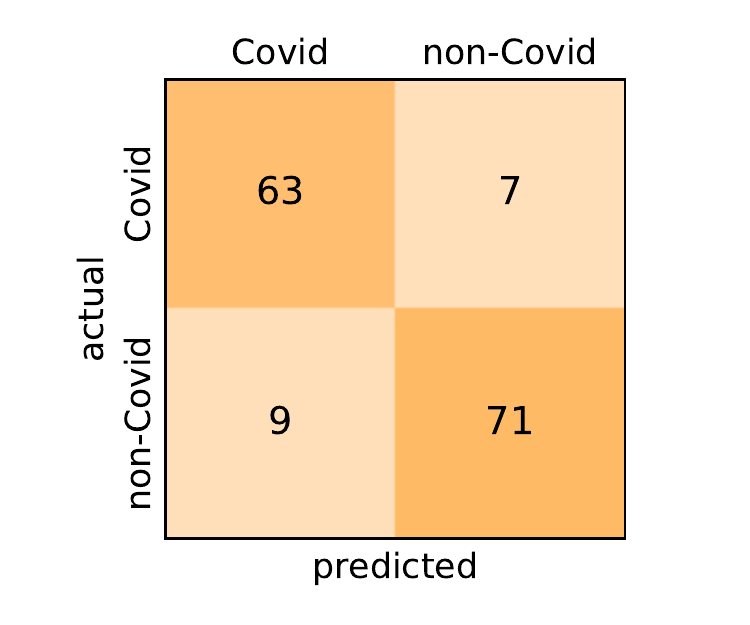}}
	\end{subfigure}		
	\begin{subfigure}[ResNeXt50]
		{\includegraphics[width=5 cm,height=4 cm]{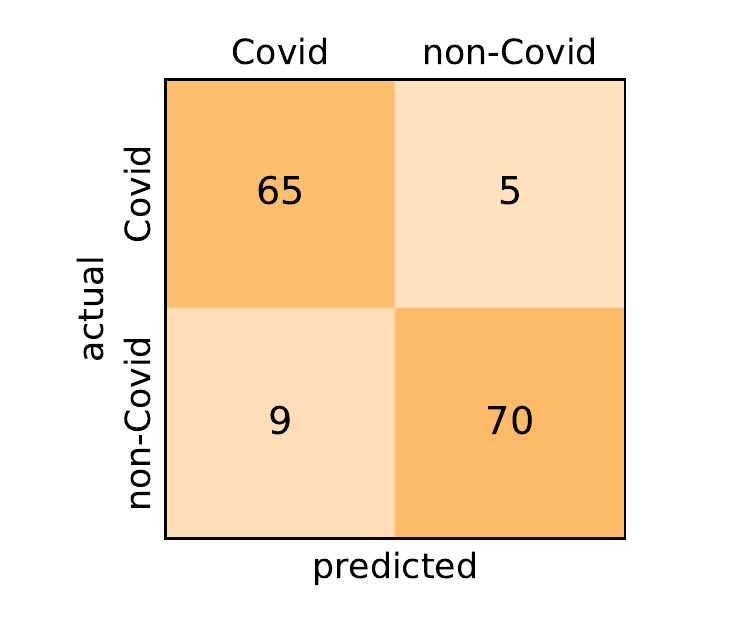}}
	\end{subfigure}	
	\begin{subfigure}[ResNeXt101]
		{\includegraphics[width=5 cm,height=4 cm]{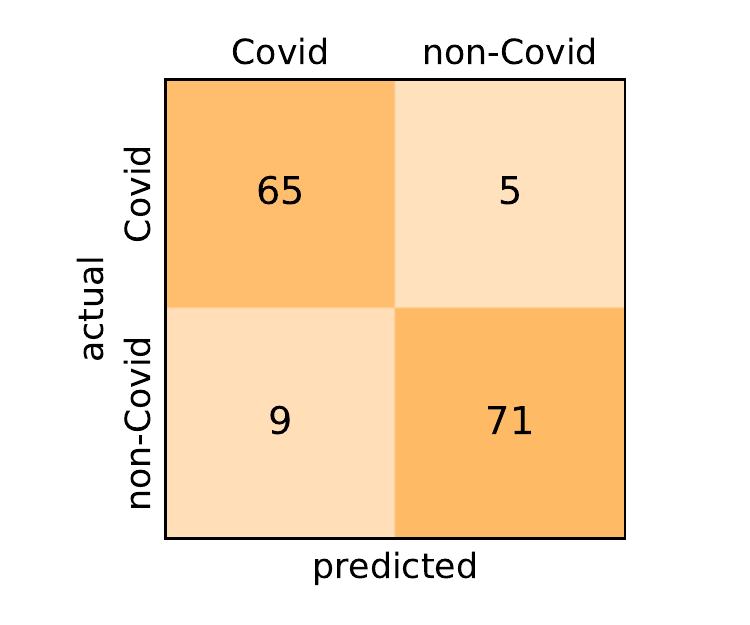}}
	\end{subfigure}		
	\begin{subfigure}[InceptionV3]
		{\includegraphics[width=5 cm,height=4 cm]{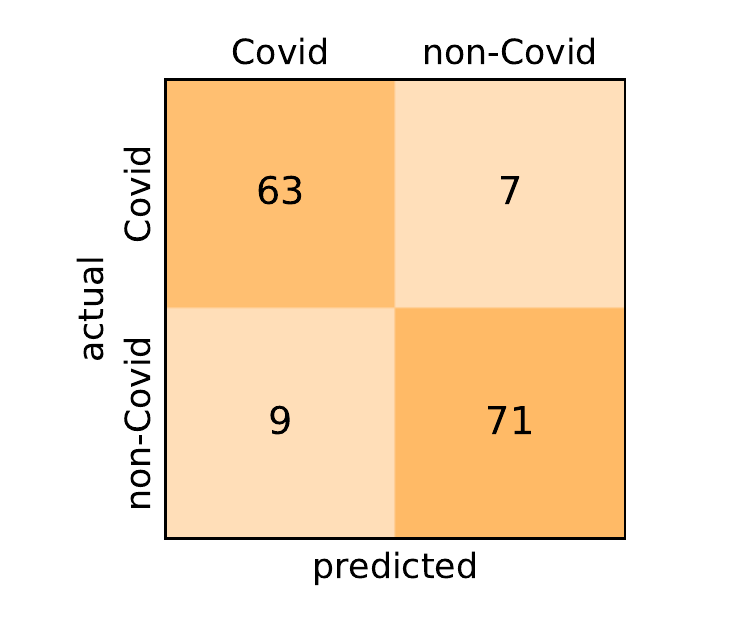}}
	\end{subfigure}
	\begin{subfigure}[Xception]
		{\includegraphics[width=5 cm,height=4 cm]{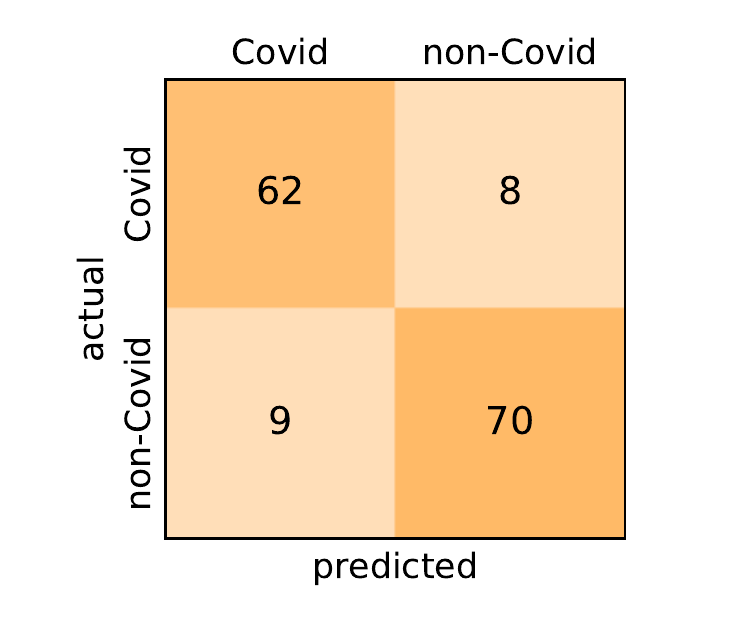}}
	\end{subfigure}	
	\begin{subfigure}[DenseNet121]
		{\includegraphics[width=5 cm,height=4 cm]{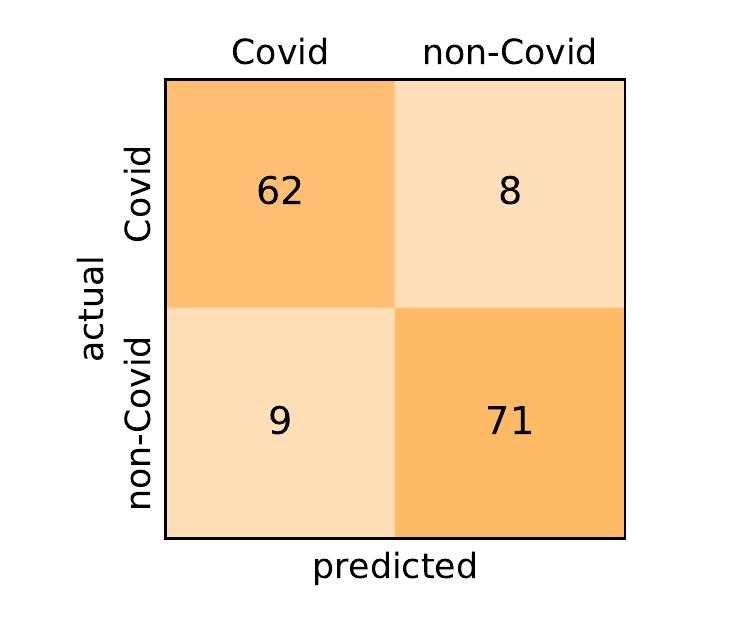}}
	\end{subfigure}	
	\begin{subfigure}[DenseNet169]
		{\includegraphics[width=5 cm,height=4 cm]{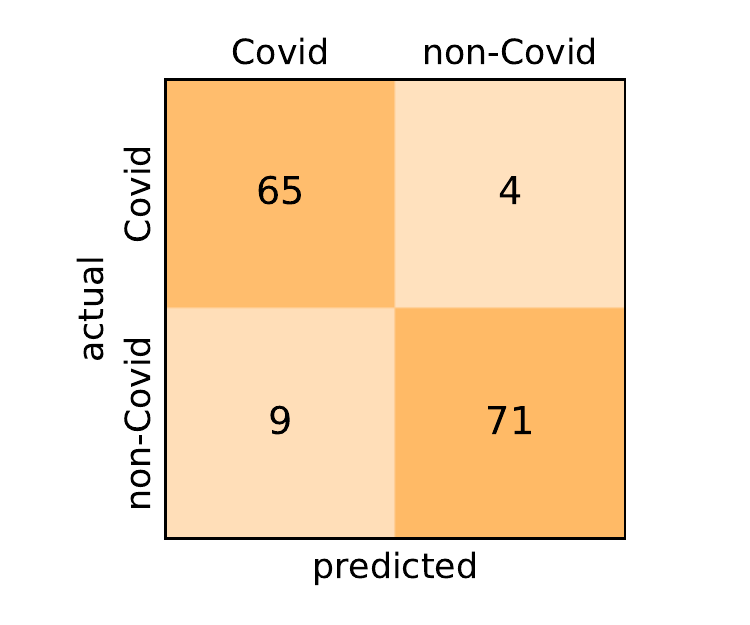}}
	\end{subfigure}	
	\begin{subfigure}[DenseNet201]
		{\includegraphics[width=5 cm,height=4 cm]{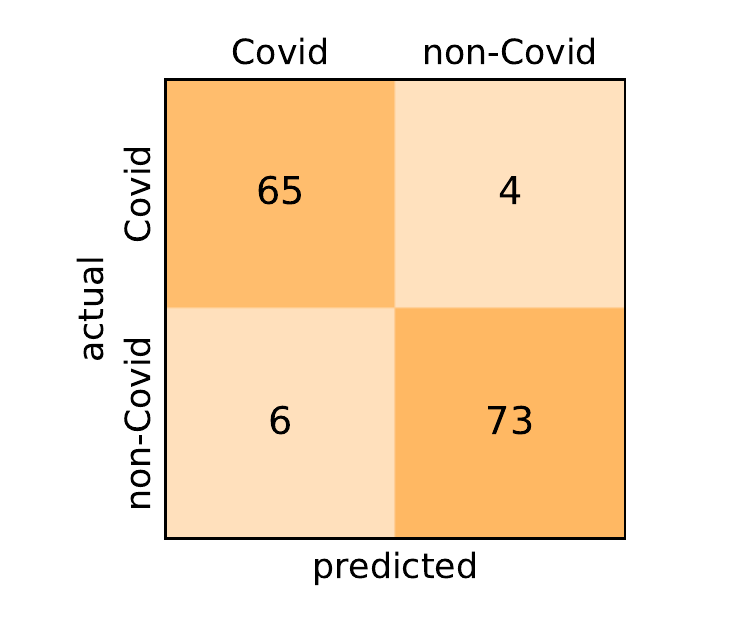}}
	\end{subfigure}	
	\caption{Confusion matrices for the different deep CNN models. These results are the average counts of the five models obtained by 5-fold cross-validation on the COVID19-CT dataset.}
	\label{confusion_matrices_ct} 
\end{figure*}

Our models achieve fairly good performance compared with the recently published work using the exact network architectures. This can bet attributed to a better optimization of our models and the effectiveness of our fine-tuning strategy using custom-sized inputs determined specifically for each architecture. Here, we see that DenseNet201 outperforms all other architectures. The model achieves average accuracy and sensitivity scores of $92.9\%$ and $93.7\%$, respectively. It also identifies all COVID-19 images with only four images, on the average, are falsely predicted as Non-COVID-19. DenseNet169 achieves the second best average accuracy of $91.6\%$ and a very high sensitivity identical to the best model. The DenseNet121 and Xception models have nearly identical results for all evaluation metrics. We observe that small-sized networks such as ResNet18 achieves comparable results with other deeper models. The SqueezeNet and ShuffleNet models perform at a similar level of accuracy. The variants of the ResNeXt models have comparable results and perform as good as the different ResNet variants. A detailed analysis on the class-wise results for individual models is presented in the confusion matrices in Figure~\ref{confusion_matrices_ct}.

\vspace{-0.5 cm}
\subsection{Visual Explanations}
\label{visualization}
This subsection provides visual explanations to make our models more transparent. We start with a 2D projection of the learned features using t-SNE \cite{maaten2008visualizing}, and then present the localization maps for highlighting the COVID-19 associated regions using Grad-CAM \cite{selvaraju2017grad}. 
\subsubsection{The t-SNE visualization}
\label{t-sne_visualization}
\begin{figure*}[!h]
	\centering
	{\includegraphics[width= 0.55\linewidth]{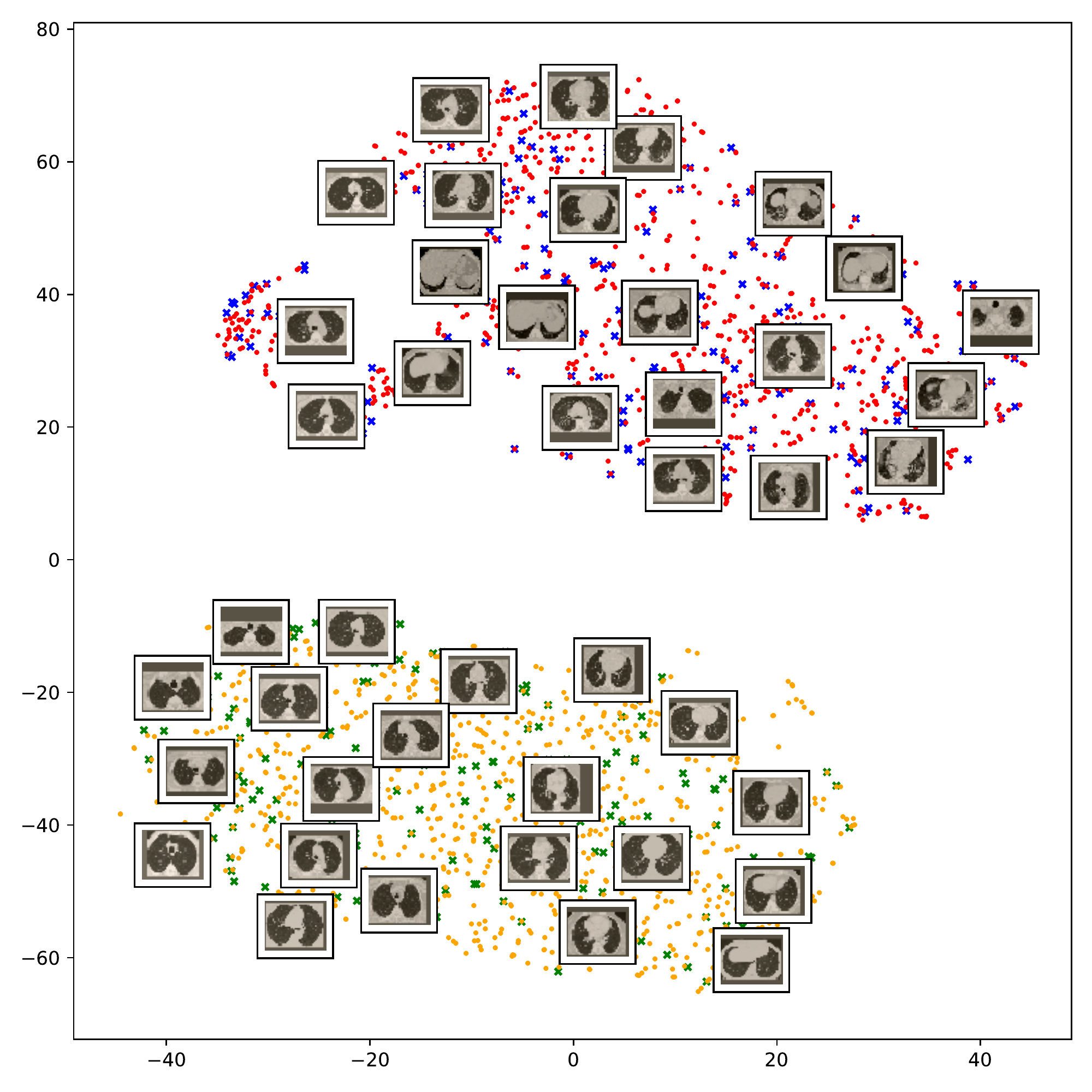}}
	\caption{Visualization of the t-SNE embeddings for the entire SARS-CoV-2 CT dataset. We clearly see two different clusters representing COVID-19 (red for train and blue for test samples) and Non-COVID-19 (yellow for train and green for test samples) classes.}
	\label{tsne_sars_cov2_dataset} 
\end{figure*}

To understand how the deep neural networks represent the CT images in the high-dimensional feature space we apply the t-SNE algorithm to visualize these features. For each image in the SARS-CoV-2 dataset we first extract the 2048-dimensional feature vector from the penultimate layer of the Inception V3 model. Next, we apply t-SNE to map the features on to 2D space and then visualize the embeddings of training and test representations. Figure~\ref{tsne_sars_cov2_dataset} clearly shows two well-separated clusters of the CT images of COVID-19 and Non-COVID-19. This indicates that the distribution of training and test features are quite similar to each other, which indicates good generalization capabilities of our model. The clear and wide margin between the two classes shows how nicely the CT images are separated in feature space.

\begin{figure*}[!h]
	\centering
	{\includegraphics[width= 0.55\linewidth]{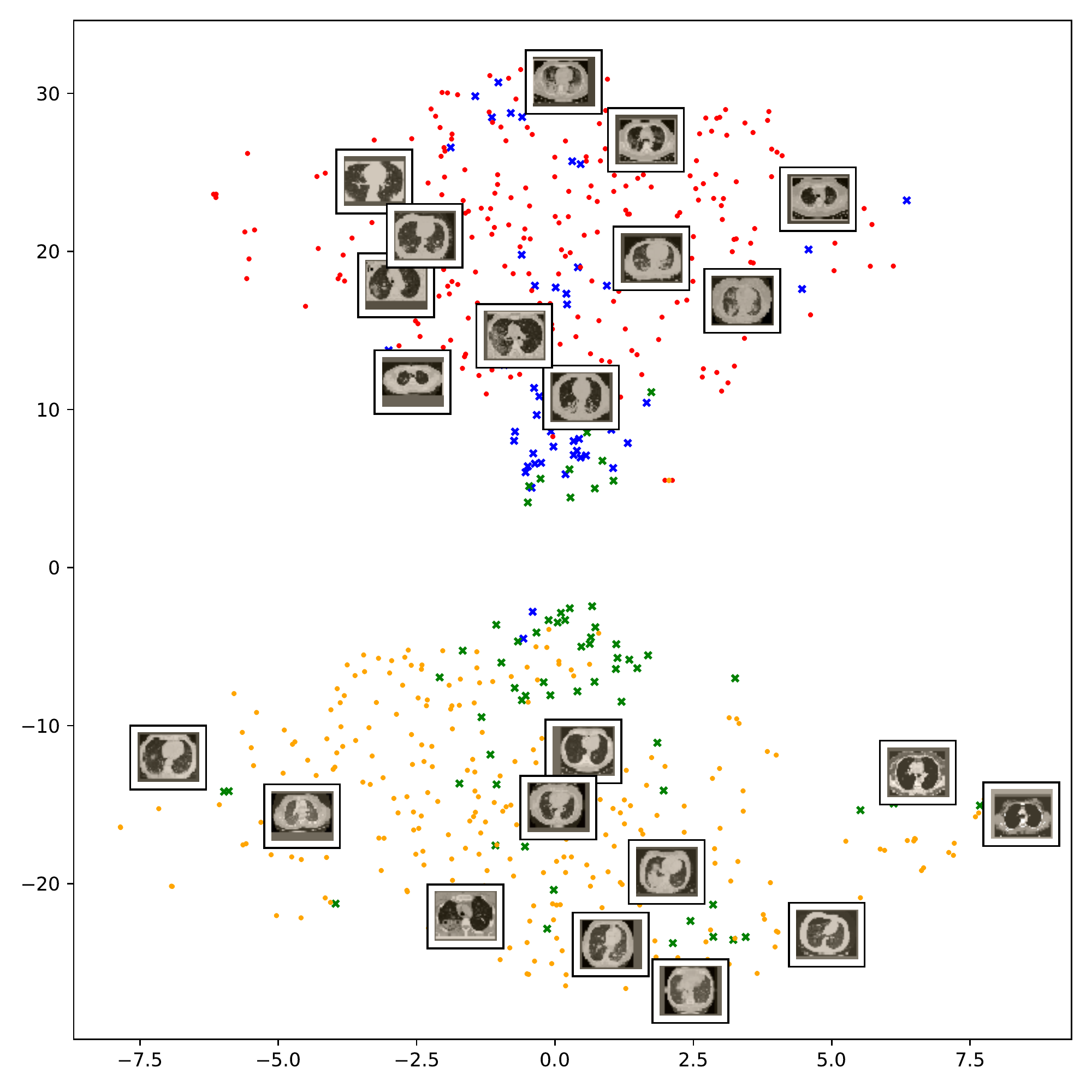}}
	\caption{Visualization of the t-SNE embeddings for the entire COVID-19 CT dataset. As in Figure~\ref{tsne_sars_cov2_dataset} we can see two different clusters representing COVID-19 and Non-COVID-19 classes.}
	\label{tsne_covid19_ct_dataset} 
\end{figure*}

We also repeat the same procedure for the COVID19-CT dataset. The feature vectors are extracted from the penultimate layer of the DenseNet169 model. The length of the feature vectors is 1664 dimensions. We again apply t-SNE to map the features on to 2D space to explore and visualize them. Figure~\ref{tsne_covid19_ct_dataset} shows two clusters representing CT images for the COVID-19 and Non-COVID-19 classes. Even though the classes are fairly distinguishable with a clear decision boundary, however, we can see that some CT images are misclassified, and more specifically the Non-COVID-19 CT images from the test set. 

\subsubsection{The Grad-CAM visualization}
\label{Grad-CAM_visualization}
\begin{figure*}[tp]
	\centering
	\begin{subfigure}
		{\includegraphics[width=6 cm]{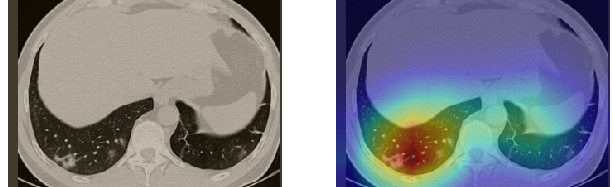}}
	\end{subfigure}	
	\begin{subfigure}
		{\includegraphics[width=6 cm]{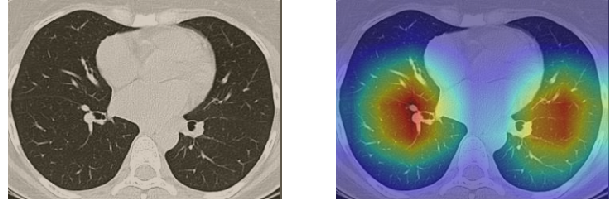}}
	\end{subfigure}		
	\begin{subfigure}
		{\includegraphics[width=6 cm]{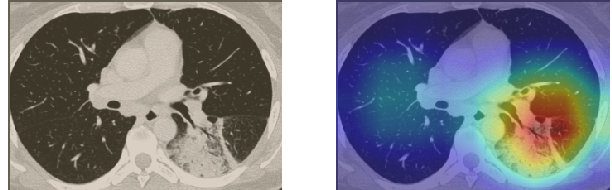}}
	\end{subfigure}
	\begin{subfigure}
		{\includegraphics[width=6 cm]{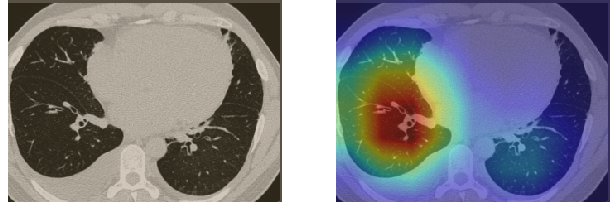}}
	\end{subfigure}	
	\begin{subfigure}
		{\includegraphics[width=6 cm]{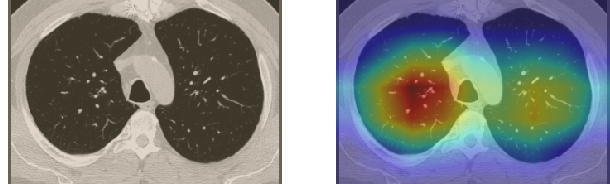}}
	\end{subfigure}	
	\begin{subfigure}
		{\includegraphics[width=6 cm]{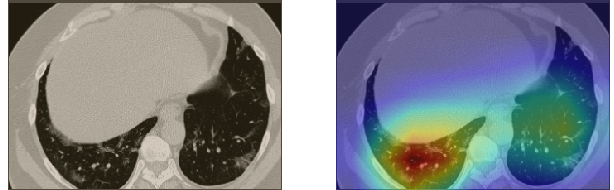}}
	\end{subfigure}		
	\begin{subfigure}
		{\includegraphics[width=6 cm]{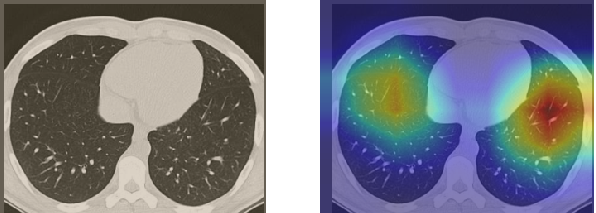}}
	\end{subfigure}
	\begin{subfigure}
		{\includegraphics[width=6 cm]{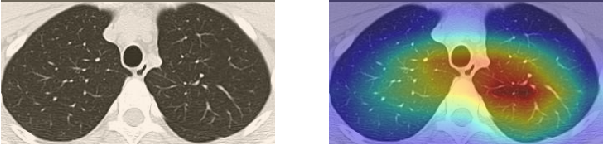}}
	\end{subfigure}
	\caption{Grad-CAM visualizations for samples CT images from the SARS-CoV-2 dataset. The InceptionV3 model correctly classified them as COVID-19 and localized the most relevant regions used for its decision. The first and third columns show CT images with COVID-19 findings, whereas the second and fourth columns represent their corresponding localization maps generated by Grad-CAM.}
	\label{grad-cam_sars_ct_images} 
\end{figure*}

In order to make our models more transparent and provide detailed visual analysis, we present the Grad-CAM localization maps obtained by different models. We consider CT images with COVID-19 abnormalities from the test set of each dataset and highlight the important regions considered for the prediction. For the SARS-CoV-2 dataset we use the Inception V3 model. Figure~\ref{grad-cam_sars_ct_images} shows the original CT images and their localization maps. Our model is capable to detect regions that show abnormalities in the CT scans. 

In a similar way, we consider classifying the test CT scans from the COVID-19 dataset by the DenseNet169 model and highlight the important regions considered for predictions. We present the original CT images and their localization maps in Figure~\ref{grad-cam_sars_ct_images}. We can also see that our model is capable to detect the COVID-19 related regions as marked (small square in some images) by expert radiologists. 

\begin{figure*}[!h]
	\centering
	\begin{subfigure}
		{\includegraphics[width=6 cm]{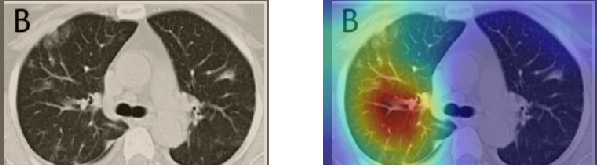}}
	\end{subfigure}	
	\begin{subfigure}
		{\includegraphics[width=6 cm]{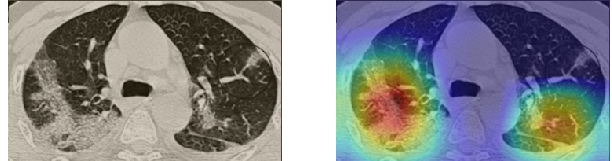}}
	\end{subfigure}		
	\begin{subfigure}
		{\includegraphics[width=6 cm]{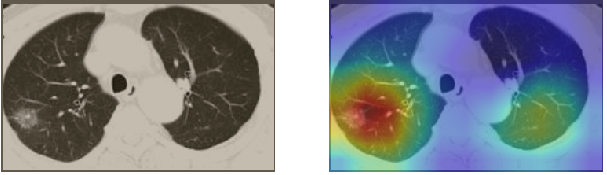}}
	\end{subfigure}
	\begin{subfigure}
		{\includegraphics[width=6 cm]{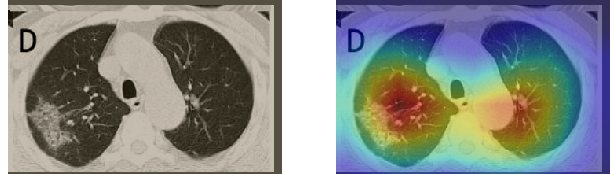}}
	\end{subfigure}	
	\begin{subfigure}
		{\includegraphics[width=6 cm]{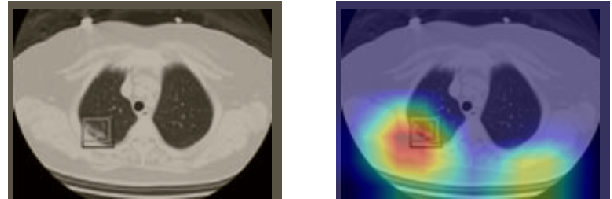}}
	\end{subfigure}	
	\begin{subfigure}
		{\includegraphics[width=6 cm]{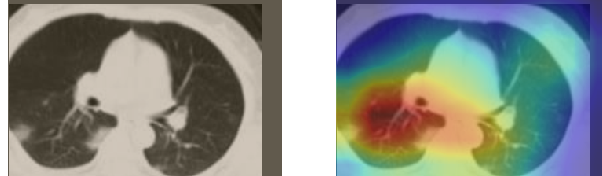}}
	\end{subfigure}		
	\begin{subfigure}
		{\includegraphics[width=6 cm]{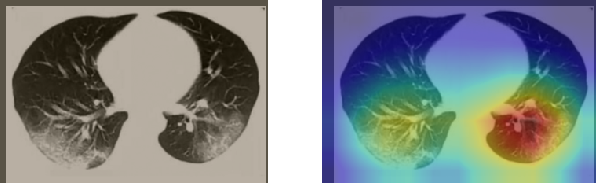}}
	\end{subfigure}
	\begin{subfigure}
		{\includegraphics[width=6 cm]{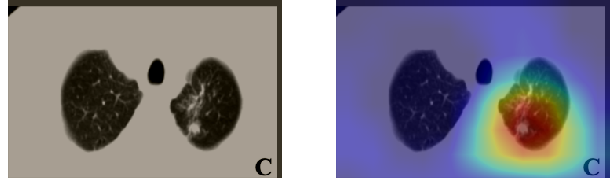}}
	\end{subfigure}
	\caption{Grad-CAM visualizations for samples CT images from the COVID19-CT dataset. The DenseNet169 model correctly classified them as COVID-19 and localized the most relevant regions as shown in the localization maps.}
	\label{grad-cam_covid19_ct_images} 
\end{figure*}

\begin{figure*}[!h]
	\centering
	\begin{subfigure}
		{\includegraphics[width=6 cm]{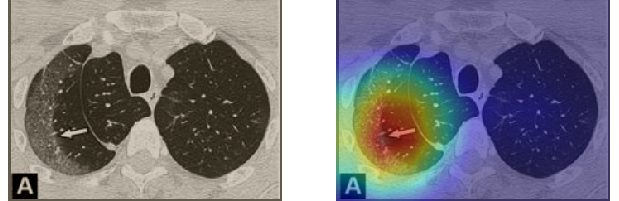}}
	\end{subfigure}	
	\begin{subfigure}
		{\includegraphics[width=6 cm]{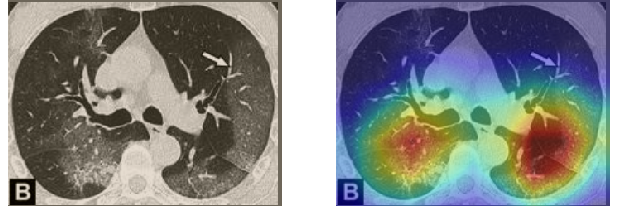}}
	\end{subfigure}		
	\begin{subfigure}
		{\includegraphics[width=6 cm]{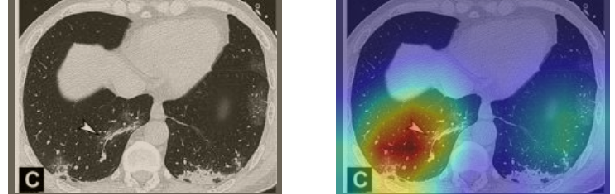}}
	\end{subfigure}
	\begin{subfigure}
		{\includegraphics[width=6 cm]{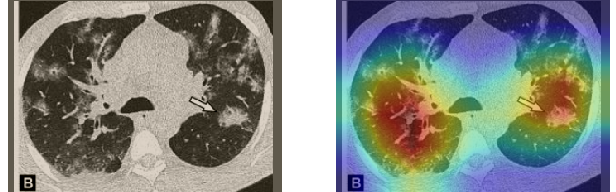}}
	\end{subfigure}	
		\begin{subfigure}
			{\includegraphics[width=6 cm]{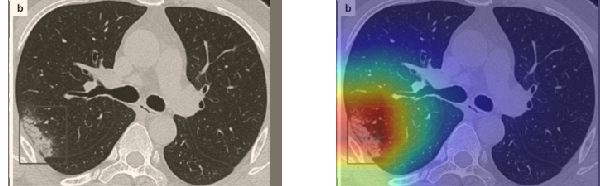}}
		\end{subfigure}	
		\begin{subfigure}
			{\includegraphics[width=6 cm]{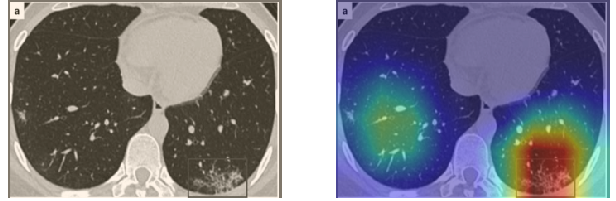}}
		\end{subfigure}		
		\begin{subfigure}
			{\includegraphics[width=6 cm]{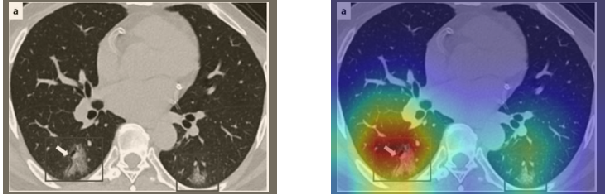}}
		\end{subfigure}
		\begin{subfigure}
			{\includegraphics[width=6 cm]{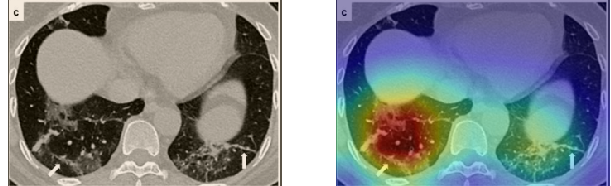}}
		\end{subfigure}	
	\begin{subfigure}
		{\includegraphics[width=6 cm]{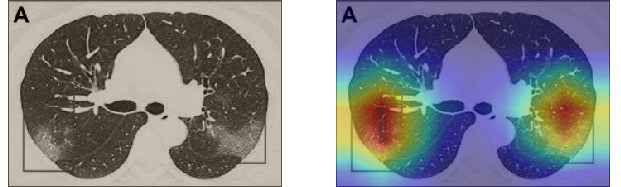}}
	\end{subfigure}	
	\begin{subfigure}
		{\includegraphics[width=6 cm]{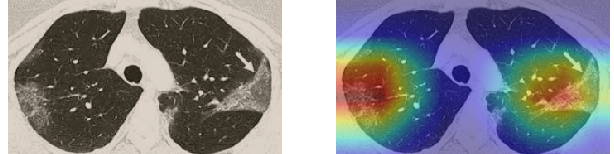}}
	\end{subfigure}		
	\begin{subfigure}
		{\includegraphics[width=6 cm]{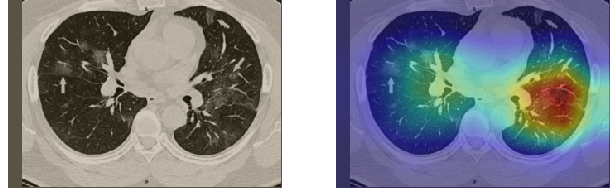}}
	\end{subfigure}
	\begin{subfigure}
		{\includegraphics[width=6 cm]{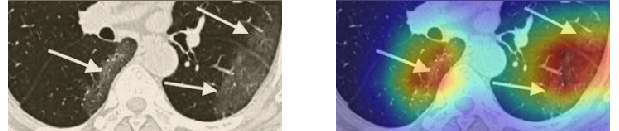}}
	\end{subfigure}
	\caption{Grad-CAM visualizations for CT images taken from two publications~\cite{ye2020chest, hani2020covid}. The CT images were correctly classified as COVID-19 and the disease-related regions are accurately localized as marked by specialists.}
	\label{external_ct_images} 
\end{figure*}

A wide variety of typical and atypical CT abnormalities have been reported for COVID-19 patients in various studies~\cite{ye2020chest, hani2020covid}. So, we tested our models on external CT images extracted from these two publications as they feature typical findings of COVID-19 pneumonia marked by specialists. In order to make sure that not any of the extracted images are unintentionally included in our datasets, specifically the COVID19-CT dataset, we use the model trained on the SARS-CoV-2 dataset. First, the InceptionV3 model is employed to classify the extracted CT images. The model is able to correctly classify the given CT images as COVID-19. Second, in order to interpret the model's generalization capabilities, we apply the Grad-CAM technique to visualize the regions of abnormalities that are considered. By assessing the different CT images in Figure~\ref{external_ct_images}, we can see that the model accurately localizes the disease-related regions. Even more interesting is the fact that the model ignores any specific marks in the images like letters and only localizes the COVID-19 related regions. These visual explanations show the success of our models to learn relevant, generic visual features related to COVID-19 and are capable to correctly classify CT images outside the datasets on which they are trained.

\begin{figure*}[!h]
	\centering
	\begin{subfigure}
		{\includegraphics[width=2 cm]{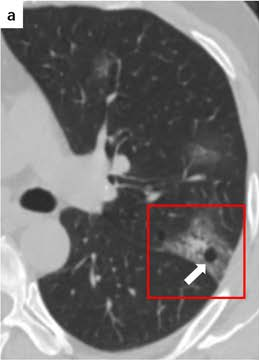}}
	\end{subfigure}	
	\begin{subfigure}
		{\includegraphics[width=2 cm]{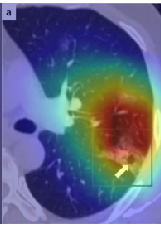}}
	\end{subfigure}		
	\begin{subfigure}
		{\includegraphics[width=2 cm]{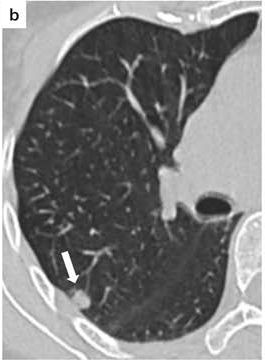}}
	\end{subfigure}
	\begin{subfigure}
		{\includegraphics[width=2 cm]{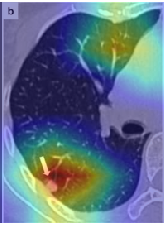}}
	\end{subfigure}	
	\begin{subfigure}
		{\includegraphics[width=2 cm]{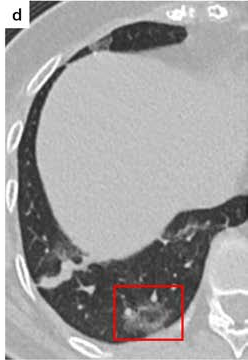}}
	\end{subfigure}	
	\begin{subfigure}
		{\includegraphics[width=2 cm]{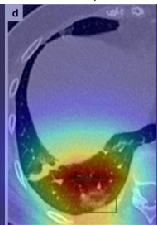}}
	\end{subfigure}		
	\caption{Grad-CAM visualizations for CT images taken from~\cite{ye2020chest}. The CT scans show different manifestations of COVID-19 marked by red frames or white arrows. Our model was able to identify them as COVID-19 and accurately localize the COVID-19 associated abnormalities.}
	\label{external_lateral_ct_images} 
\end{figure*}

Figure~\ref{external_lateral_ct_images} shows various CT scans where only one lung is visible. The CT scans are also extracted from the paper~\cite{ye2020chest} and show different CT manifestations of COVID-19 pneumonia marked by red squares. The InceptionV3 model is capable to classify them correctly as COVID-19, although it is trained on CT scans where the entire lung is visible. Intriguingly, when applying Grad-CAM we can see that all regions of abnormalities are accurately localized. This also proves the potential of our model to detect COVID-19 abnormalities in CT images outside the dataset used for training.

\vspace{-0.5 cm}
\section{Conclusion}
\label{conclusion}
We proposed different deep learning based approaches for accurate COVID-19 detection using chest CT images. The most advanced deep network architectures and their variants were considered and extensive experiments were conducted on the two datasets with the largest amount of CT images available so far. Moreover, we investigated different configurations and determined custom-sized input for each network to achieve the best detection performance. The resulting networks showed a significantly improved performance for detecting COVID-19. Our models achieved state-of-the-art performance with an average accuracy of $99.4\%$ and $92.9\%$, and a sensitivity score of $99.8\%$ and $93.7\%$ on the SARS-CoV-2 CT and COVID19-CT datasets, respectively. This indicates the effectiveness of our proposed approaches and the potential of using deep learning for fully automated and fast diagnosis of COVID-19. In order to explain the obtained results we employed two visualization methods. First, we explored the learned features using the t-SNE algorithm and the resulting visualizations showed well-separated clusters for COVID-19 and Non-COVID-19 cases. We also assessed the obtained networks using the Grad-CAM algorithm to obtain high-resolution visualizations showing the discriminative regions of abnormalities in the CT images. Moreover, we tested our models on external CT images from different publications. Our models were capable to detect all COVID-19 cases and accurately localize the COVID-19 associated regions as marked by expert radiologists. 

\begin{acknowledgements}
The authors gratefully acknowledge the constructive feedback from Prof. Dr. J{\"o}rg Barkhausen from the Clinic for Radiology and Nuclear Medicine at the Universit{\"a}tsklinikum Schleswig-Holstein (UKSH), L{\"u}beck. The work of Hammam Alshazly was supported by the Bundesministerium f{\"u}r Bildung und Forschung (BMBF) through the KI-Lab Project. The work of Christoph Linse was supported by the Bundesministeriums f{\"u}r Wirtschaft und Energie (BMWi) through the Mittelstand 4.0-Kompetenzzentrum Kiel Project. 
\end{acknowledgements}


\bibliographystyle{IEEEtran}      
\bibliography{references}

\end{document}